# FILM: Mapping organellar metabolism by mid-infrared photothermal modulated fluorescence

Jianpeng Ao[1, 2, 3#], Jiaze Yin[1, 2#], Haonan Lin[1, 2, 3], Guangrui Ding[1, 2], Youchen Guan[4], Marzia Savini[4], Bethany Weinberg[3, 6], Dashan Dong[1, 2], Qing Xia[1, 2], Zhongyue Guo[2, 5], Bowen Liu[1], Biwen Gao[7], Ji-Xin Cheng[1, 2, 3, 5, 6, 7*], Meng C. Wang[4*]

[#] Contributed equally

[1] Department of Electrical and Computer Engineering, Boston University, Boston, MA 02215

[2] Photonics Center, Boston University, Boston, MA 02215

[3] HHMI Janelia Visiting Scholar Program, Ashburn, VA 20147

[4] HHMI Janelia Research Campus, Ashburn, VA 20147

[5] Department of Biomedical Engineering, Boston University, Boston, MA 02215

[6] Graduate Program in Molecular Biology, Cell Biology, & Biochemistry, Boston University, Boston, MA 02215

[7] Department of Chemistry, Boston University, Boston, MA 02215

* Corresponding Authors: Ji-Xin Cheng jxcheng@bu.edu; Meng C. Wang mengwang@janelia.hhmi.org

## Abstract

Metabolism unfolds within specific organelles in eukaryotic cells. Lysosomes are highly metabolically active organelles, and their metabolic states dynamically influence signal transduction, cellular homeostasis, and organismal physiopathology. Despite the importance of lysosomal metabolism, a method for its *in vivo* measurement is currently lacking. Here, we report a Fluorescence-detected mid-Infrared photothermaL Microscope (FILM) implemented with optical boxcar demodulation, AI-assisted data denoising, and spectral deconvolution, to map metabolic activity and composition of individual lysosomes in living cells and organisms. Using this method, we uncovered lipolysis and proteolysis heterogeneity across lysosomes within the same cell, as well as early-onset lysosomal dysfunction during organismal aging. Additionally, we discovered organelle-level metabolic changes associated with diverse lysosomal storage diseases. This method holds the broad potential to profile metabolic fingerprints of individual organelles within their native context and quantitatively assess their dynamic changes under different physiological and pathological conditions, providing a high-resolution chemical cellular atlas.

# Main

## Introduction

Metabolism is essential for biological systems to sustain their physiological activities, and its dysfunction contributes to diverse diseases[1, 2]. In multicellular organisms, metabolic processes are compartmentalized across multiple levels, from organelles to tissues. Therefore, understanding the spatial organization of metabolism is crucial for both biological and biomedical research. However, this remains technically challenging, particularly at the scale of organelles. Organelles are fundamental structural and functional units within eukaryotic cells, each specializing in distinct metabolic processes. Organelle-specific immunoprecipitation has facilitated the enrichment of specific organelles from different tissues for mass-spectrometry based metabolic profiling [3-6]. However, spatial information within the organelles' native cellular context is lost during this process. On the other hand, microscopic imaging of fluorescence-labelled organelles has revealed their structural organization, dynamics and heterogeneity *in vivo* [7-9]; however, even with specific metabolite sensors, it provides limited insight into their metabolic complexity [10-12].

Infrared (IR) absorption spectroscopy simultaneously fingerprint a wide range of molecules based on their vibrational signatures[13,14]. Mid-infrared photothermal (MIP) microscopy, which measures the photothermal effects caused by IR absorption, has further advanced IR spectroscopic imaging to submicron resolution[15-17], enabling chemical imaging of biomolecules in living cells[18-20]. Leveraging thermal sensitivity of fluorescent reporters[21, 22], Fluorescence-detected mid-Infrared photothermaL Microscopy (F-MIP, termed as FILM here) enables organelle-level imaging of certain molecules[23-25]. However, the reported work[23, 24] suffers from substantial photobleaching and requires an exposure time three orders of magnitude longer than conventional fluorescence microscopy to capture a full fingerprint spectrum. This limitation makes it nearly impossible to comprehensively map metabolic activity *in vivo* across a broad spectral range and substantially hinders in-depth mapping of organellar complexity in relation to their metabolic states.

Here, we developed an optical boxcar demodulation scheme, together with a synchronized IR-visible laser scanner and an artificial intelligence (AI)-assisted self-supervised hyperspectral denoiser, to simultaneously enhance the photothermal signal, reduce the fluorescence exposure time, and mitigate the photobleaching issue. The FILM signal is followed by an unmixing algorithm to quantify biomolecular contents. This upgraded FILM system enables hyperspectral imaging of organelles in the entire fingerprint window (1000 to 1800 $cm^{-1}$). We have applied this system for *in vivo* metabolic profiling of lysosomes—organelles that play vital roles in nutrient sensing, macromolecular recycling, signaling transduction, diseases, and aging[26-29], revealing their distinctive metabolic fingerprints and changes during physiological aging and under various disease conditions.

# Results

## Fluorescence-detected mid-Infrared photothermaL Microscope (FILM)

FILM bridges fluorescence emission with IR absorption by probing photothermal changes across a broad range of IR wavelengths that encode the chemical signatures of biomolecules. Mid-infrared photons excite molecular vibrational modes, which subsequently relax into heat (**Fig. 1a**). The resulting local temperature increase enhances nonradiative relaxation in nearby fluorescent probes, reducing their quantum yield for fluorescence emission via dynamic quench. This modulation in fluorescence intensity serves as an indicator of IR absorption.

In the previous point-scan system [23, 24], a continuous-wave (CW) visible laser is used to excite fluorescent molecules (**Fig. 1b**). The fluorescence intensity changes are then demodulated using a lock-in amplifier (LIA) at the IR repetition rate. However, due to the low duty-cycle nature of the pulsed photothermal process, most fluorescence photons contribute to shot noise rather than photothermal signal. More importantly, prolonged CW exposure leads to substantial photobleaching. To overcome these limitations, we implemented frequency-demodulated optical boxcar detection to eliminate photons that do not contribute to the photothermal signal, thereby reducing photobleaching (**Fig. 1c**). In addition to demodulation at the IR pump frequency, a pair of pulsed visible probes, synchronized with IR excitation, was used to selectively gate emission events at the peak of the temperature rise, referred to as the "hot" state, and the subsequent cooled "cold" state. The system is depicted in **Fig. 1d**. High-speed laser scanning geometry was implemented at 30 microseconds per pixel to further minimize photobleaching. A multichannel pulse generator synchronized the IR laser at a repetition rate $f_{IR}$ of 200 kHz and the visible laser $f_{vis}$ at 400 kHz. Fluorescence was detected using a silicon photomultiplier (SiPM), and the signal was directly demodulated at $f_{IR}$ using LIA.

We first evaluated photobleaching reduction by modulating the duty cycle of the visible excitation while maintaining constant peak power (**Fig. 1e**). After 100 frames of scanning, pulsed excitation at 40% and 20% duty cycle reduced photobleaching to 29.3% and 16.6% of the CW condition, respectively (**Fig. 1e**). We then assessed signal intensity as a function of duty cycle and observed a "stable zone" in which the signal remained nearly unchanged when the duty cycle was reduced from 100% (i.e., CW) to 30% (**Fig. 1f**). We therefore selected a 30% duty cycle, corresponding to a 750 ns pulse duration, which preserved the signal while reducing photobleaching. Overall, optical-boxcar enhanced FILM reduces fluorescence excitation time by over 100 times compared to the previously reported point-scan system[23, 24] (**Supplementary Fig. 1**).

Additionally, the pulsed excitation light served as a $2f$ carrier, shifting high odd-order harmonic signals into the demodulation frequency, thereby enhancing signal amplitude relative to CW condition, while maintaining the same average power (**Extended Data Fig. 1**, **Supplementary Fig. 2**; **Supplementary Note 1**). The two gating windows of the excitation light used in optical boxcar inherently function as a time-resolved measurement, making it less sensitive to slow heat diffusion background that universally exists in water environment (**Extended Data Fig. 2**; **Supplementary Note 2**).

Using this system, we performed hyperspectral FILM on Rhodamine 6G-labelled *S. aureus* (**Fig. 1g**), acquiring 160 frames spanning 980-1780 $cm^{-1}$. The extracted spectrum resolved characteristic peaks from nucleic acids and protein Amide II and Amide I bands (**Fig. 1h**). Spectral fidelity was further validated by comparing FILM spectra of bacteria with scattering-based MIP spectra, as well as FILM spectra of LysoSensor DND-189 stained DMSO with reference ATR-FTIR spectra (**Extended Data Fig. 3**), confirming reliable fingerprinting of fluorescently labelled objects.

## Hyperspectral FILM imaging of lysosomes and AI assisted data analysis

To image lysosomes with optical-boxcar enhanced FILM, we identified a lysosome-specific thermosensitive dye, LysoSensor DND-189, whose fluorescence intensity decreased by 12% upon a 10 Kelvin temperature increase (**Supplementary Fig. 3**). Lysosomes in intestinal cells of wide-type *C. elegans* were labeled with this dye and imaged to capture lysosome-specific hyperspectral datasets (**Supplementary Video 1**). As the pixel integration time was reduced to 30 microseconds to avoid photobleaching, the photothermal signal exhibited a relatively low signal-to-noise ratio (SNR).

To recover SNR without increasing integration time, we harnessed a self-supervised deep learning denoising algorithm, Self-permutation Noise2Noise Denoising (SPEND)[30]. SPEND generates two image stacks from a single low-SNR hyperspectral dataset by permuting the hyperspectral stack along the ω dimension into odd and even slices that were alternately concatenated to form independent measurements of the same field of view (FOV) (**Fig. 2a**). A 3D U-Net is then trained using these paired stacks as input and target, enabling effective learning of noise statistics and object priors while avoiding information leakage between adjacent pixels due to point-scan imaging (**Supplementary Fig. 4**; **Supplementary Note 3**). Once trained, the model can be applied in batch to denoise hyperspectral datasets acquired under same conditions.

Lysosomes were visible in the raw FILM image at 1711 $cm^{-1}$ but disappeared at 1797 $cm^{-1}$ (**Fig. 2b**), confirming chemical selectivity of FILM, which is further supported by the complete set of wavenumber frames in **Supplementary Fig. 5**. Intensity profiles along the marked line revealed that lysosomes with weaker signals were nearly obscured by noise fluctuations in raw images (**Fig. 2c**). After SPEND processing, noise in non-lysosomal regions was effectively suppressed, enabling clear visualization of weak lysosomal signal (**Fig. 2b, c; Supplementary Video 1**), which was further validated by hyperspectral stack projection and DC fluorescence images (**Supplementary Fig. 6**). Noise reduction was also evident in the spectral dimension, where SPEND yielded smoother single-lysosome hyperspectral profiles and reduced frame-to-frame fluctuations (**Fig. 2d**). Quantitatively, SPEND improved image SNR by 26.9× and the spectral SNR by 5.3× (**Fig. 2e**; **Supplementary Note 4**), outperforming alternative denoising methods for analyzing FILM data (**Extended Data Fig. 4**).

After baseline correction, power normalization, and spectral internal normalization, fingerprint spectra of individual lysosomes were extracted. While ratiometric analysis of specific chemical bands was feasible for qualitative visualization, quantitative decomposition of multiple biomolecular contents required spectral unmixing (**Fig. 2a**). We first constructed a spectral matrix from calibrated lysosomal spectra and performed least absolute shrinkage and selection operator (LASSO) [31, 32] based unmixing using eight reference spectra acquired from pure standards on the same system (**Fig. 2a**; **Supplementary Fig. 7**; **Supplementary Table 1**). Reconstruction of lysosomal spectra using LASSO alone revealed noticeable mismatches relative to the measured spectra (**Fig. 2f**), reflecting the high chemical complexity of lysosomes, which contain hundreds of molecular species[33].

To improve the unmixing performance, we introduced Multivariate Curve Resolution (MCR)[34] prior to LASSO to refine the reference spectra using lysosomal data. An augmented MCR strategy incorporating initial reference spectra was implemented to stabilize spectral updates and maintain physical interpretability. The resulting reconstructed spectra showed improved agreement with lysosomal spectra (**Fig. 2g**). Quantitative evaluation using cosine similarity and Euclidean distance confirmed superior spectral fitting with MCR-LASSO compared to LASSO alone (**Fig. 2h**; **Supplementary Note 5**). Together, these AI-driven analyses enabled us to quantitatively measure biomolecular contents within individual lysosomes and compare them between conditions.

## FILM reveals hydrolytic heterogeneity of lysosomes

Using the AI-assisted FILM system, we imaged lysosomes in live *C. elegans* (**Supplementary Video 2**). Fluorescence imaging localized individual lysosomes (**Fig. 3a**), while hyperspectral FILM revealed a previously inaccessible chemical dimension of these organelles through IR spectra (**Fig. 3b**). Lysosomes exhibited spectral features distinct from surrounding regions visualized by autofluorescence (**Fig. 3b**; **Supplementary Fig. 8**; **Supplementary Note 6**). Spectral phasor analysis further enabled robust segmentation of lysosomes from their surroundings (**Supplementary Fig. 9**). Strikingly, the spectra varied among different lysosomes (**Fig. 3b**), indicating a highly heterogeneous lysosomal population even within the same cell. Control experiments comparing lysosomal spectra with that of the dye itself, together with dye concentration estimates based on fluorescence intensity, confirmed that the observed spectral features did not originate from the thermosensitive probe (**Supplementary Fig. 10**).

Comparison of lysosomal and surrounding-region spectra revealed two characteristic lysosomal peaks around 1587 $cm^{-1}$ and 1711 $cm^{-1}$. By examining FILM spectra of standard mixtures, including proteins, amino acids (AA), triglycerides (TAG, lipid ester) and free fatty acids (FFA), we tentatively assigned these features to AA and FFA (**Supplementary Fig. 8c**; **Supplementary Note 7**). To independently validate these assignments, we inactivated CTNS-1 transporter that exports cysteine from the lysosome[35], led to increased AA accumulation and produced a modest but statistically notable enhancement at 1587 $cm^{-1}$ ($P = 0.013$, two-sided two-sample t-tests; **Extended Data Fig. 5a**). Additionally, overexpression of LIPL-4 lipase that hydrolyzes triglyceride into FFA[36], resulted in a pronounced increase at 1711 $cm^{-1}$ ($P = 2.14 \times 10^{-52}$, two-sided two-sample t-tests; **Extended Data Fig. 5b**). These perturbations provided biological support for the assignment of the 1587 $cm^{-1}$ and 1711 $cm^{-1}$ feature to AA and FFA, respectively. Our results revealed that the lysosomal spectrum exhibits a higher presence of AA and FFA, which are consistent with the active hydrolytic function of lysosomes[37, 38] and thus support the ability of FILM to specifically profile the metabolic composition of lysosomes *in vivo*.

Building on these assignments, we defined the ratio of 1587 $cm^{-1}$ and 1649 $cm^{-1}$ (AA/protein) as a proxy for proteolytic activity and the ratio of 1711 $cm^{-1}$ and 1741 $cm^{-1}$ (FFA/lipid esters) as a proxy for lipolytic activity, leveraging the distinct IR signatures of macromolecules and their degradation products. Consistent with prior lysosome-targeted lipidomics studies showing enhanced lipolysis in LIPL-4 overexpression (*lipl-4 Tg*) worms[36], FILM imaging revealed an increased 1711/1741 $cm^{-1}$ ratio in lysosomes of *lipl-4 Tg* worms ($P = 3.11 \times 10^{-18}$, two-sided two-sample t-tests; **Extended Data Fig. 5c**), supporting this ratio as a qualitative readout for lysosomal lipolysis *in vivo*. Pixel-wise ratio maps and parallel-set visualizations further highlighted stronger hydrolytic activities in lysosomes compared to surrounding regions (**Extended Data Fig. 6**).

More importantly, lysosomes exhibiting high proteolytic activity did not fully overlap with those showing high lipolytic activity (**Fig. 3c**), suggesting metabolic heterogeneity within the lysosomal population. Based on the two activity ratios, lysosomes were categorized into three groups: high proteolytic activity, high lipolytic activity, and high activity in both (**Fig. 3d**). Correlation analysis shows only weak relationships between hydrolytic activity and lysosomal size (**Extended Data Fig. 7a**, **b**), as well as activity and fluorescence intensity (**Extended Data Fig. 7c**, **d**), indicating that the observed metabolic signatures were not driven by probe concentration or size effects. This metabolic heterogeneity was also detected in mammalian lysosomes by FILM (**Fig. 3e** and **Extended Data Fig. 8**). These results define the metabolic landscape of lysosomes in wild-type cells and reveal the functional basis of lysosomal heterogeneity.

In addition to intracellular heterogeneity, we observed substantial intercellular heterogeneity in lysosomal metabolism under normal physiological conditions (**Extended Data Fig. 9**). Using data from day-2-old wild-type worms, the t-SNE embedding consistently separated lysosomal spectra into three functional subpopulations, consistent with the classification shown in **Fig. 3**. These results indicate that the distribution of metabolically distinct lysosomal subpopulations varies markedly across cells.

## FILM tracks lysosomal metabolic changes during aging

Metabolic dysfunction is a hallmark of aging[2]. To investigate age-related metabolic changes at the organellar level in lysosomes, we generated the ratiometric images of 1587/1649 $cm^{-1}$ (proteolytic activity), as well as 1711/1741 $cm^{-1}$ (lipolytic activity) in *C. elegans* at adult Day 2, Day 4, Day 6, and Day 10 (**Fig. 4a**) with corresponding quantitative analysis shown in **Fig. 4b**. Notably, both hydrolytic activities declined with age (**Fig. 4b**), with the decrease occurring early in life, as early as Day 4 of adulthood, prior to the onset of aging-related mortality, indicating early-onset lysosomal metabolic dysfunction during aging.

To capture age-dependent changes across the entire fingerprint spectrum, we extracted dozens of lysosomal spectra from each age group and generated heatmaps (**Fig. 4c-d)**. Clear spectral differences were observed across ages. To further highlight these differences, we performed z-score analysis relative to the total average spectrum (**Fig. 4e**). Lysosomes from Day 2 animals exhibited enriched AA and FFA signatures, whereas those from Day 4 showed stronger Amide I and Amide II features. Lysosomes from Day 6 and Day 10 animals displayed more prominent features in the lower wavenumber region (1060–1350 $cm^{-1}$) that can be attributed to nucleic acids and carbohydrates.

Dimensionality reduction using t-SNE revealed that lysosomal spectra from Day 2 formed a compact and well-separated cluster from those of later ages (**Fig. 4f**). Day 4 spectra also clustered distinctly from Day 6 and Day 10, whereas the latter two showed greater overlap (**Fig. 4f**). Euclidean distance analysis confirmed that spectra from each age group were most similar within their own group, with the smallest inter-group difference observed between Day 6 and Day 10 (**Supplementary Fig. 11**), indicating relatively minor spectral divergence at late ages. In contrast, intra-group variability increased from Day 4 compared to Day 2 (**Fig. 4g**), suggesting that lysosomal metabolic heterogeneity becomes more pronounced with aging.

To quantitatively interpret these spectral changes, we performed spectral decomposition using eight reference components, including protein, AA, FFA, TAG, ceramides, glycogens, deoxyribonucleic acid (DNA), and cholesterol ester (CE). We found that average levels of macromolecules, including DNA, ceramides, triglycerides, and glycogens, increase with age (**Fig. 4h**), suggesting their age-related accumulation within lysosomes.

## FILM profiles metabolic changes associated with LSDs

Metabolic dysfunction of lysosomes underlies lysosomal storage diseases (LSD), leading to the accumulation of undegraded macromolecules within lysosomes[27]. To date, it remains challenging to assess metabolic changes at the lysosomal level under those pathological conditions. We hypothesized that FILM provides an avenue to address this challenge. To test this hypothesis, we knocked down several well-conserved LSD genes using RNA interference (RNAi) in *C. elegans*.

We first fingerprinted the lysosomes of five Day 2 RNAi groups together with their controls, whose lysosomal profiles represent the normal metabolic state (**Fig. 5a-b**). Heatmap visualization revealed pronounced alterations in lysosomal metabolic composition following RNAi-mediated inactivation of each LSD gene (**Fig. 5a**). Consistent with the results shown in **Fig. 4d**, control lysosomes exhibited dominant peaks near 1587 $cm^{-1}$ and 1711 $cm^{-1}$, corresponding to AA and FFA, respectively. Compared with controls, *nuc-1* RNAi led to a substantial increase in spectral intensity near 1100 $cm^{-1}$ and 1294 $cm^{-1}$, with milder increases observed in other RNAi conditions (**Fig. 5b**). In the 1530-1730 $cm^{-1}$ range, all RNAi groups displayed altered peak shapes and relative intensities at 1587, 1649, and 1711 $cm^{-1}$.

To quantitatively interpret these changes, we decomposed the lysosomal spectra to resolve eight chemical components (**Fig. 5c**). RNAi inactivation of *nuc-1* and *asah-2* led to the accumulation of DNA, ceramides and TAG in lysosomes, while the level of FFA decreased. *aagr-2* RNAi increased ceramides, TAG and glycogens content. *lipl-3* and *ncr-1* RNAi caused accumulation of TAG, glycogens, while FFA levels decreased. In addition to the negative correlation between protein and AA (Pearson's r = -0.74) and between TAG and FFA (Pearson's r = -0.65), consistent with proteolysis and lipolysis, FFA and DNA also exhibit a negative correlation (Pearson's r = -0.63), suggesting that defects in DNA degradation may impact the lipolytic activity of lysosomes (**Fig. 5d**).

We next applied FILM to investigate metabolic changes of lysosomes in mammalian cells with NPC1 knockout (NPC1KO), a well-established model of Niemann-Pick disease type C characterized by lysosomal accumulation of cholesterol and glycosphingolipids[39]. Lysosomal fingerprint spectra were acquired from both wild-type (WT) and NPC1KO cells (**Fig. 5e-f**). In contrast to *C. elegans*, mammalian lysosomes were dominated by Amide I and Amide II bands, reflecting higher protein content. Compared with WT, lysosomes in the NPC1KO cells exhibited stronger signals between 1100 $cm^{-1}$ and 1250 $cm^{-1}$, as well as changes in the 1530-1730 $cm^{-1}$ range (**Fig. 5f**).

Quantitative analysis revealed that NPC1KO lysosomes exhibited reduced levels of FFA, alongside increased accumulation of DNA, ceramides, TAG, CE, and glycogens (**Fig. 5g)**, indicating globally impaired macromolecular degradation. Correlation analysis showed a positive correlation between TAG and glycogens (Pearson's r = 0.64), suggesting coordinated accumulation, whereas FFA exhibit negative correlations with DNA (Pearson's r = -0.67), ceramides (Pearson's r = -0.78) and glycogens (Pearson's r = -0.63) (**Fig. 5h**). These relationships indicate that lysosomal lipolysis may be modulated by defects in nucleic acid and sphingolipid degradation, as well as by glycogen accumulation.

## FLIM images mitochondria and lipid droplets

FILM establishes a platform for organelle-specific chemical imaging in living systems that extends beyond lysosomes. As representative demonstrations, we labeled lipid droplets in HeLa cells with Lipi-Red and mitochondria with MitoTracker Green in parallel experiments, yielding clearly distinguishable spectral signatures (**Extended Data Fig. 10**). Lipid droplets exhibited a dominant peak at 1741 $cm^{-1}$, attributable to esterified C=O stretching, consistent with their triglyceride-rich composition. In contrast, mitochondria displayed protein-dominated spectra accompanied by phosphate-associated features, likely arising from ATP and other phosphorylated metabolites related to tricarboxylic acid cycle activity. Together, these results demonstrate that FILM enables chemical profiling of diverse organelles, underscoring its generalizability as a platform for organelle-resolved metabolic imaging in living systems.

## Discussion

FILM equipped with optical boxcar demodulation, AI-based denoising, and spectral decomposition provides a technical platform for profiling organelle-level metabolism in living cells and organisms and capturing their dynamic changes under different physiological and pathological conditions. MIP microscopy with electronic boxcar detection has been implemented to extract signals from water background by harnessing the photothermal dynamics[40]. Our optical boxcar strategy, implemented with a pulsed probe beam synchronized with the IR pulses, effectively mitigates the photobleaching issues that were prevalent in early point-scan F-MIP studies[23, 24]. It also reduces the solvent background interference, particularly in aqueous environments, beyond the wide-field geometry primarily confined to dry samples[25]. Together with SPEND-associated denoising and optimized MCR-LASSO spectral decomposition, these methodological advancements collectively ensure high sensitivity, specificity, and robustness of FILM.

Vibrational microspectroscopy techniques, leveraging coherent Raman scattering or optical photothermal detection of vibrational absorption, offer powerful tools for spatial metabolic profiling[19, 41-44] with high spatial resolution but typically lack organelle specificity. FILM bridges vibrational and fluorescence imaging modalities, which provides chemical fingerprints and organellar specificity simultaneously. Unlike stimulated Raman or IR up-conversion fluorescence[45-47], which rely on the co-excitation of specific fluorophores to achieve superior sensitivity but confine chemical information to the dye itself, FILM operates as a decoupled process, where fluorescent molecules function as reporters to sense the surrounding molecules. In comparison, fluorescence-guided MIP for co-localizing vibrational imaging[48, 49] often suffers from spatial mismatches caused by focal plane shifts, and scattering-based MIP is prone to ring artifacts that are heightened by environmental solvent interference.

In this study, we chose lysosomes to demonstrate the application of FILM, given their involvement in diverse macromolecular processing and their highly dynamic metabolic activities [50, 51]. Importantly, FILM is also applicable to other metabolically active organelles, such as lipid droplets and mitochondria. Further advancements in instrumentation, such as enhancing hyperspectral acquisition speed through thermal deposition multiplexing, could further improve imaging efficiency and throughput[52]. Although MCR was used to refine references using measured data, MCR-LASSO remains a reference-guided, supervised spectral decomposition approach, with components interpreted relative to predefined standards. Accurate quantitative interpretation thus depends on contextually appropriate reference spectra, making reliable standards crucial for robust results. The quantitative findings here are based on carefully selected references but should not be seen as definitive descriptions of lysosomal composition. Additionally, incorporating analysis beyond the mid-IR fingerprint region to include CH-, NH-, and OH-stretch vibrational modes could further improve chemical specificity and provide more comprehensive insights into organelle metabolic heterogeneity in future work.

Overall, as a proof-of-principle study, this work demonstrates that FILM holds pronounced promise for advancing *in vivo* metabolic investigation across scales. This precise and versatile imaging technology will provide critical insights into cellular mechanisms underlying aging, metabolic disorders, and disease pathogenesis.

## Acknowledgements

We thank Lynne Chantranupong and Elif Ozsen (Boston University) for helpful discussions on lysosomal metabolism, Roberto Zoncu (University of California, Berkeley) for providing the HEK293 NPC1KO and HEK293 WT cell lines, and Dennis Kim and Sylvia Fischer (Boston Children's Hospital) for lending us a stereo microscope. This work was supported by NIH R35GM136223 (J.-X.C) and Howard Hughes Medical Institute (M.C.W). The authors acknowledge the Visiting Scientist Program at HHMI Janelia Research Campus.

## Author Contributions Statement

J.A., J.Y., J.-X.C. and M.C.W. conceived the project. J.A., J.Y. constructed the FILM system. M.C.W., Y.G. and M.S. prepared the *C. elegans* samples. J.A. prepared mammalian cell samples. B.W. prepared the *Shigella flexneri* sample. B.G. prepared the *S aureus* sample. J.A. collected and analyzed data. H.L. and G.D. aided the SPEND and MCR-LASSO analysis. D.D., Q.X. and Z.G. contributed to the data analysis. B.L. contributed to the standards measurement during the revision. J.A., J.Y., J.-X.C. and M.C.W. wrote the manuscript with input and approval from all authors. J.-X.C. and M.C.W supervised the project.

## Competing Interests Statement

J.-X.C. declares financial interest with Photothermal Spectroscopy Corp at Santa Barbara. Other authors declare no competing interests.

## Figure Legends

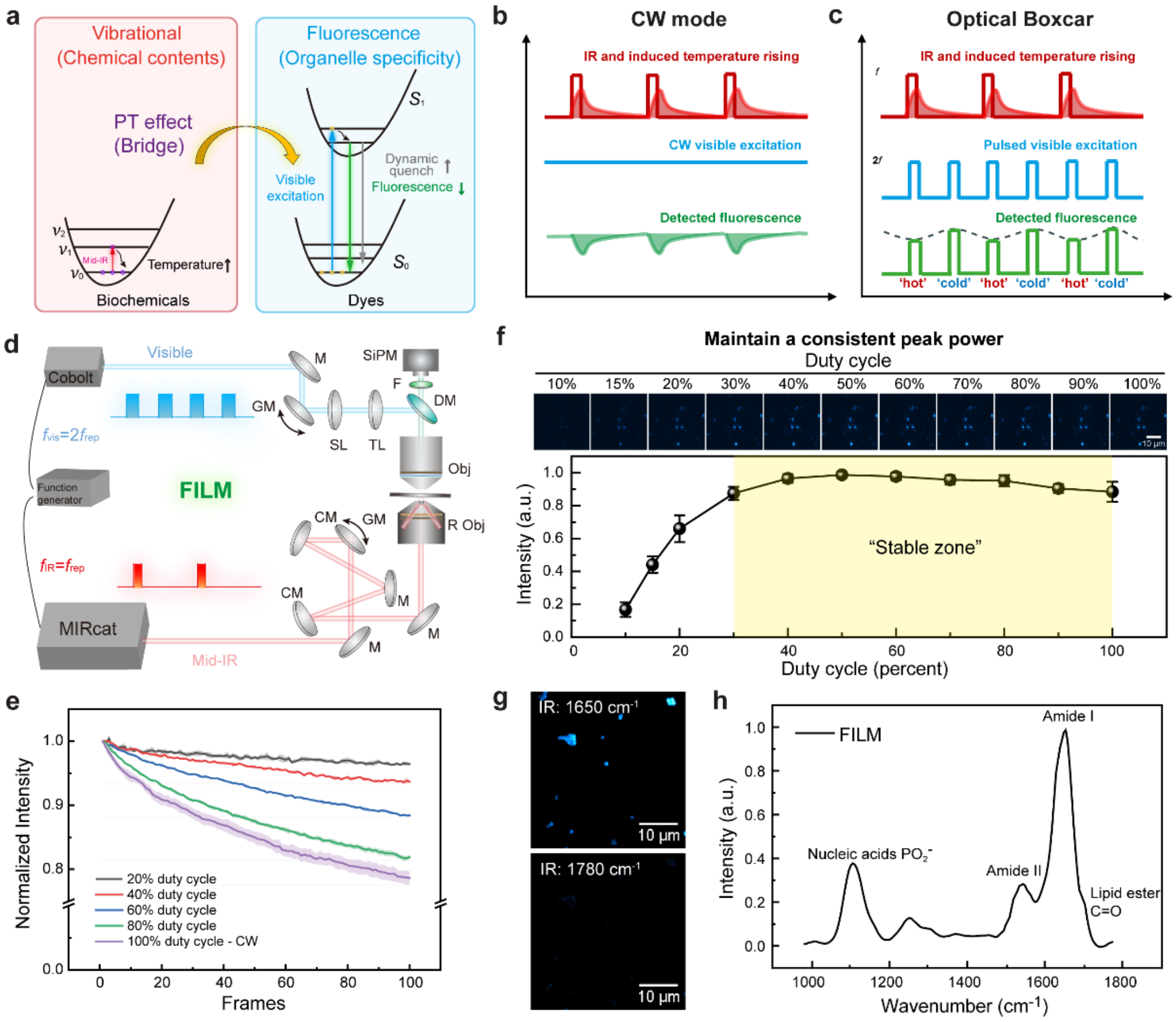

**Figure 1 Development of FILM. a,** Principle of fluorescence detected mid-infrared photothermal microscopy depicted by energy diagram. Mid-IR absorption induces vibrational energy transitions in biochemical molecules, resulting in localized photothermal (PT) effects and a temperature increase. This rise in temperature accelerates the dynamic quenching of fluorophores, leading to a reduction in fluorescence emission. **b**, The previous continuous-

wave (CW) fluorescence excitation schematic recorded the entire IR-induced PT dynamics. **c**, The optical boxcar schematic selectively recorded the 'hot' and 'cold' states to remove non-contributing photons, thereby mitigating photobleaching. **d,** Schematic of the experimental setup for the FILM microscope. M: reflection mirrors; GM: galvo mirrors; CM: concave mirrors; SL: scan lens; TL: tube lens; DM: dichroic mirror; Obj: objective; R obj: reflective objective; F: filter; SiPM: silicon photomultiplier. **e**, Photobleaching curves of standard fluorescence beads (n=3) under different excitation duty cycles. Shaded area indicates the standard deviation (s.d.) of photobleaching measurements. **f**, FILM signal of *Shigella flexneri* expressing GFP measured with different duty cycles visible light (n=5). Statistical data are presented as mean ± s.d. **g**, FILM images of *S. aureus* at 1650 $cm^{-1}$ and 1780 $cm^{-1}$. **h**, FILM spectrum of single *S. aureus*. a.u., arbitrary units. Scale bar: 10 μm.

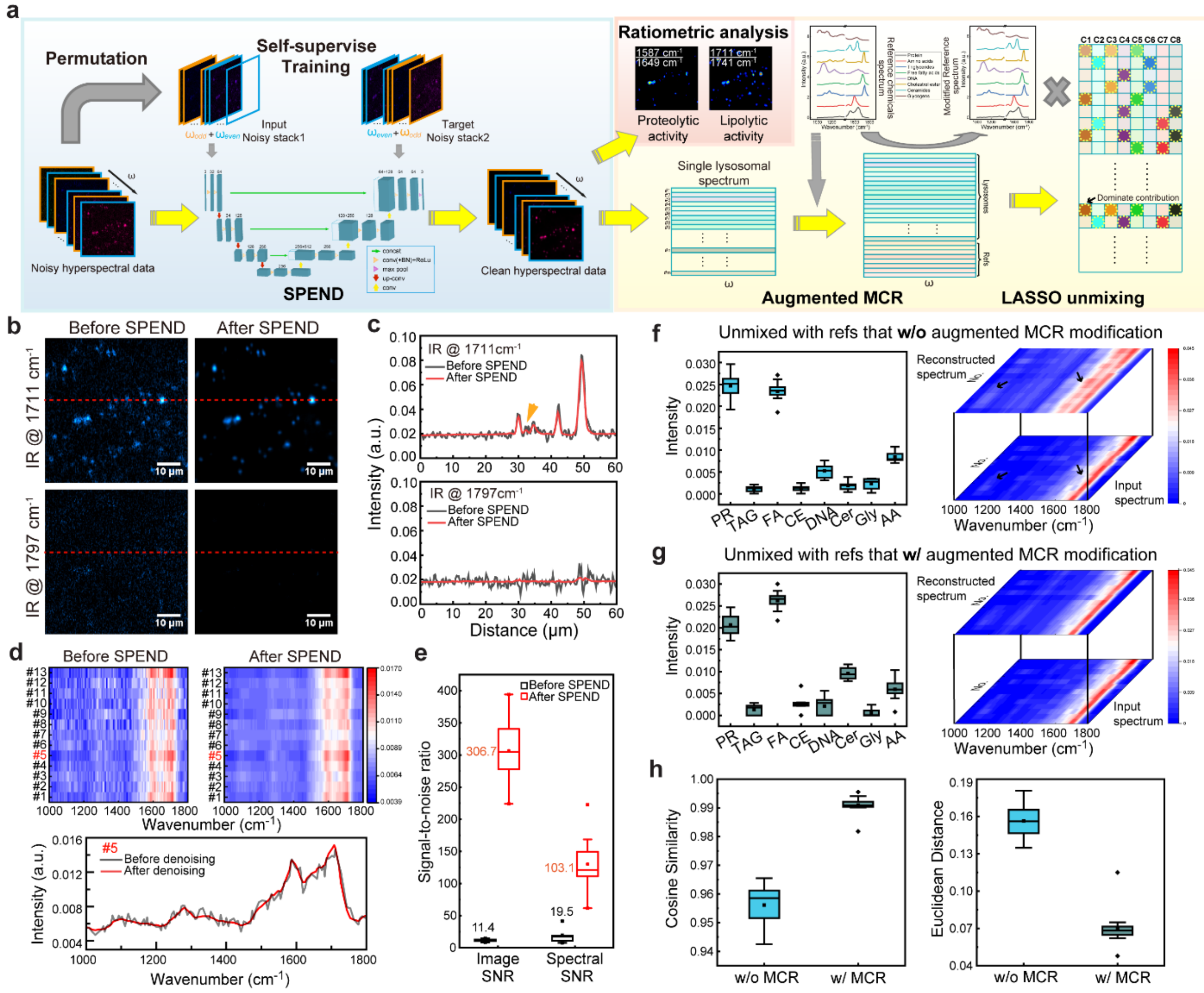

**Figure 2 AI-assisted FILM hyperspectral imaging and analysis**. **a,** Workflow of AI-assisted hyperspectral data analysis. The left panel represented a deep learning based self-supervised denoising algorithm, called Self-permutation Noise2Noise Denoising (SPEND). The raw noisy hyperspectral data were first rearranged into two different sequences with permutation process. Next, the two sets of noisy data were served as the input and target for a U-net training. The trained network was then applied to denoise raw hyperspectral data. In the schematic, concat denotes concatenate, conv(+BN)+ReLU indicates convolution followed by batch normalization and rectified linear unit activation, max pool represents max pooling, up-conv refers to up-convolution (transposed convolution), and conv denotes convolution. The right panel represented the ratiometric analysis and MCR-LASSO spectral unmixing process. Reference spectrum of pure chemicals, acquired with the same instrument, were modified with augmented MCR based on the lysosomal data and then fed to LASSO for spectral unmixing and quantification. **b**, The comparison of FILM images of lysosomes acquired with IR at 1711 $cm^{-1}$ and 1797 $cm^{-1}$ before and after SPEND denoising. **c**, Intensity profiles along the red dotted lines marked in **b**. **d,** The comparison of raw FILM spectrum without calibration before and after SPEND processing. **e**, Quantification of image SNR and spectral SNR before and after SPEND denoising (n=13). **f**, LASSO unmixing with unmodified references and comparison of original calibrated input and reconstructed spectrum (n=13). **g**, LASSO unmixing with MCR modified references and comparison of original calibrated input and reconstructed spectrum (n=13). **h**, The comparison of cosine similarity and Euclidean distance with and without augmented MCR modification (n=13). Scale bar: 10 μm. In **e**, **f**, **g**, and **h**, the boxes show the interquartile range (IQR), the centerlines indicate medians and the lines outside the boxes extend to 1.5 times the IQR.

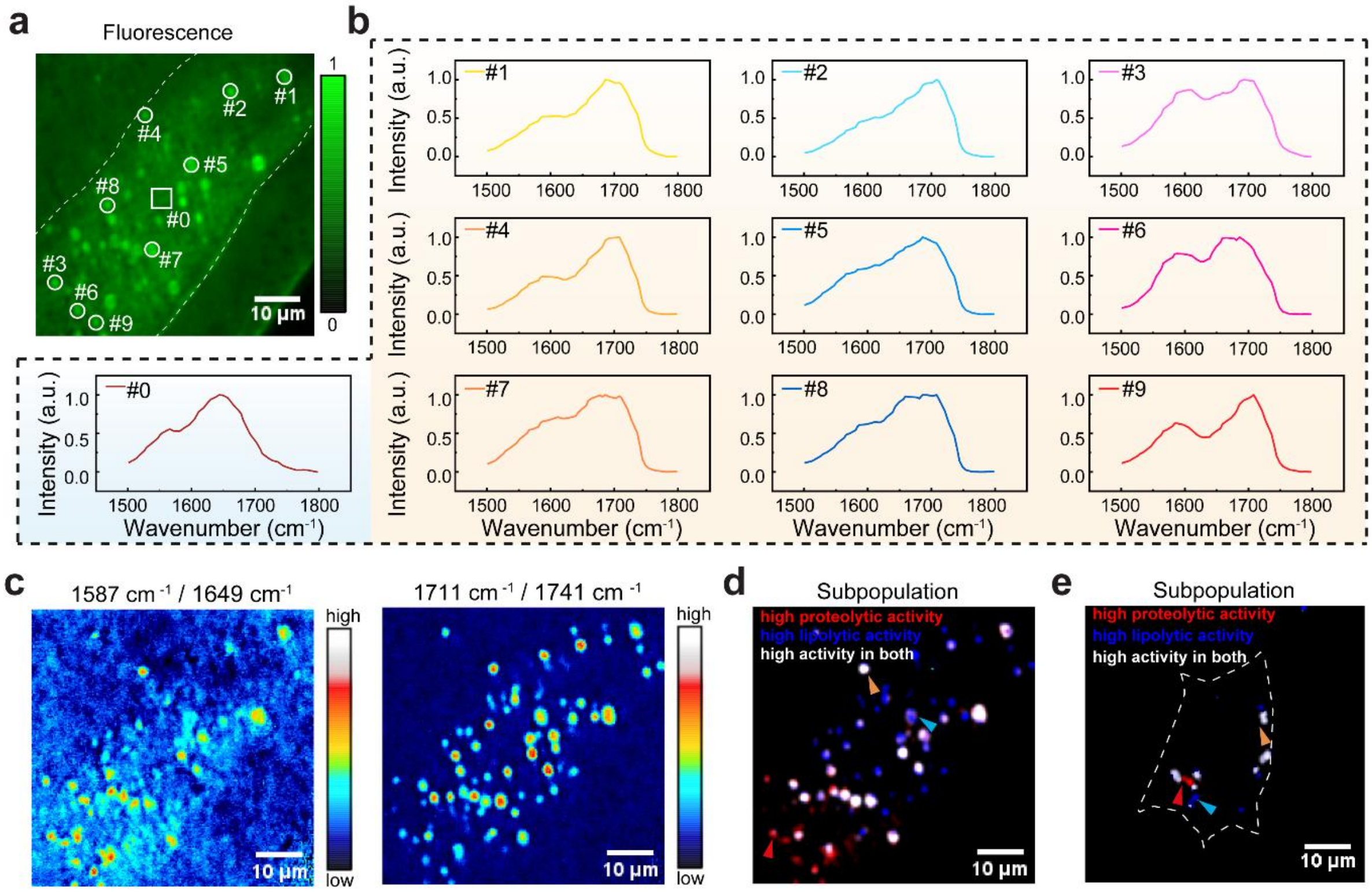

**Figure 3 Hydrolytic heterogeneity of lysosomes revealed by FILM**. **a**, Fluorescence image of *C. elegans* labelled with LysoSensor DND189. **b**, FILM spectra of individual lysosomes and surrounding region marked in **a. c**, Ratio-metric mapping of intensity ratios at 1587/1649 $cm^{-1}$ (proteolytic activity) and 1711 /1741 $cm^{-1}$ (lipolytic activity), representing proteolysis activity and lipolysis activities, respectively. **d**, Classification of lysosomal subpopulations based on the two ratios shown in **c**. **e**, Classification of lysosomal subpopulations of mammalian lysosomes. Scale bar: 10 μm. Representative results are shown from three independent experiments.

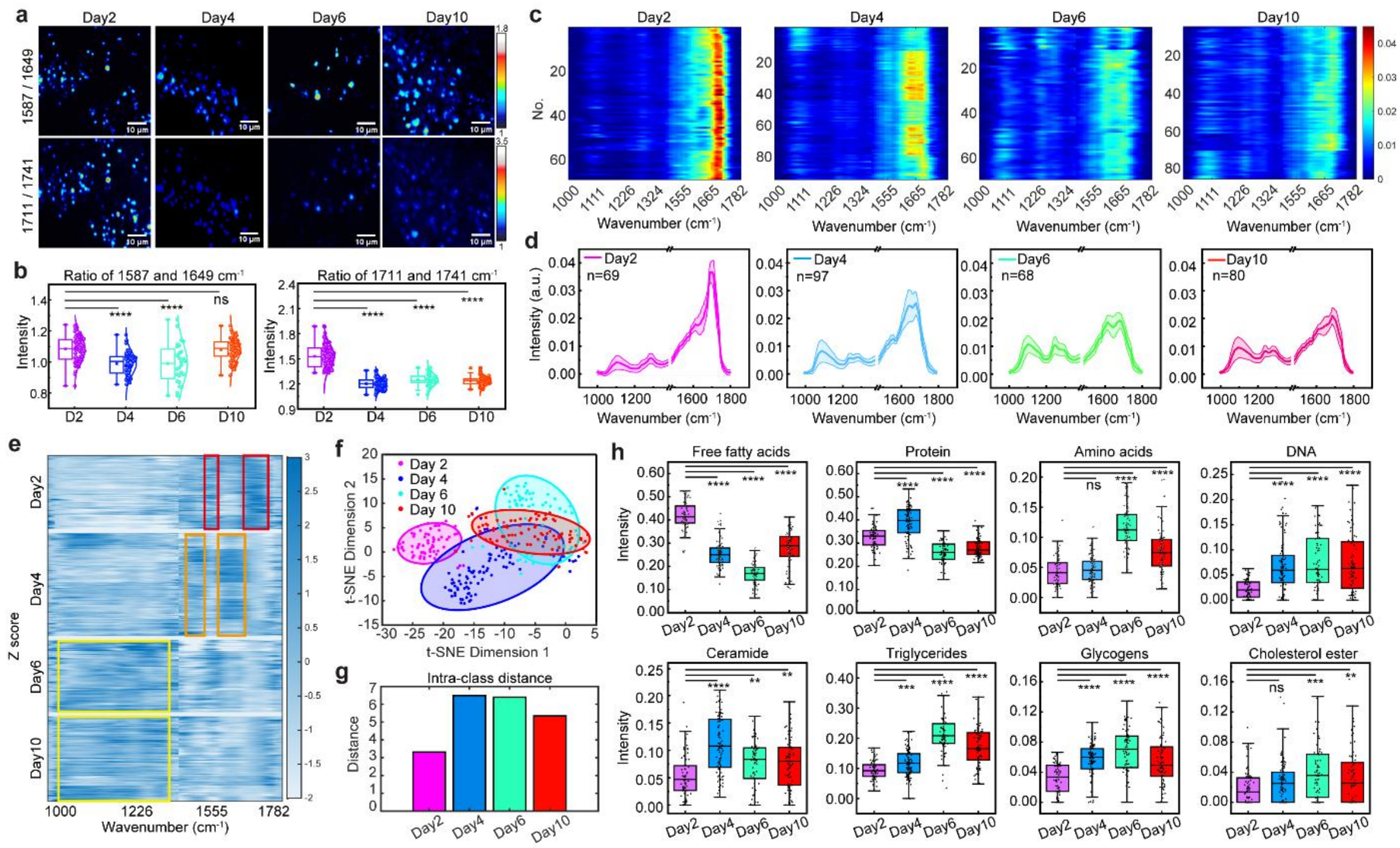

**Figure 4 Age-related metabolic changes at lysosomal scale**. **a**, Ratio-metric mapping of intensity ratios at 1587/1649 $cm^{-1}$ and 1711/1741 $cm^{-1}$ across worms of different ages. **b**, Quantitative comparison of the two intensity ratios among four age groups (Two-sample t-test comparing with Day2 group: *, $p<0.05$; **, $p<0.01$; ***, $p<0.001$; ****, $p<0.0001$; ns, not significance). **c**, Heatmap of lysosomal fingerprint spectra extracted from worms in four age groups (n=69 for Day2, n=97 for Day4, n=68 for Day6 and n=80 for Day10 derived from five to seven independent experiments), highlighting spectral variations with age. Each row represented a lysosomal spectrum. **d**, Representative average spectra for each age group, showing age-dependent metabolic differences. Shaded area indicates the standard deviation. **e**, Z-score heatmap of different age groups. Red boxes highlight signal regions with the higher intensity for Day2 group. Orange boxes indicate signal regions with the higher intensity for Day4 group. Yellow boxes highlight signal regions with the higher intensity for Day6 and Day10 groups. **f**, t-SNE visualization of all spectra, displaying clustering patterns based on age-related spectral features. Each dot indicates a lysosomal spectrum. Shaded area indicates 85% confidence interval. **g**, Intra-cluster distance analysis from t-SNE, where larger distances indicate poorer clustering and greater heterogeneity within the data. **h**, High-content analysis of metabolic profiles across the four age groups, identifying age-related trends. All comparisons were made relative to Day 2 group, whose lysosomal profiles are representative of a normal metabolic state. (Two-sample t-test: *, $p<0.05$; **, $p<0.01$; ***, $p<0.001$; ****, $p<0.0001$; ns, not significance). Scale bar: 10 μm. In **b** and **h**, the boxes show the interquartile

range (IQR), the centerlines indicate medians and the lines outside the boxes extend to 1.5 times the IQR.

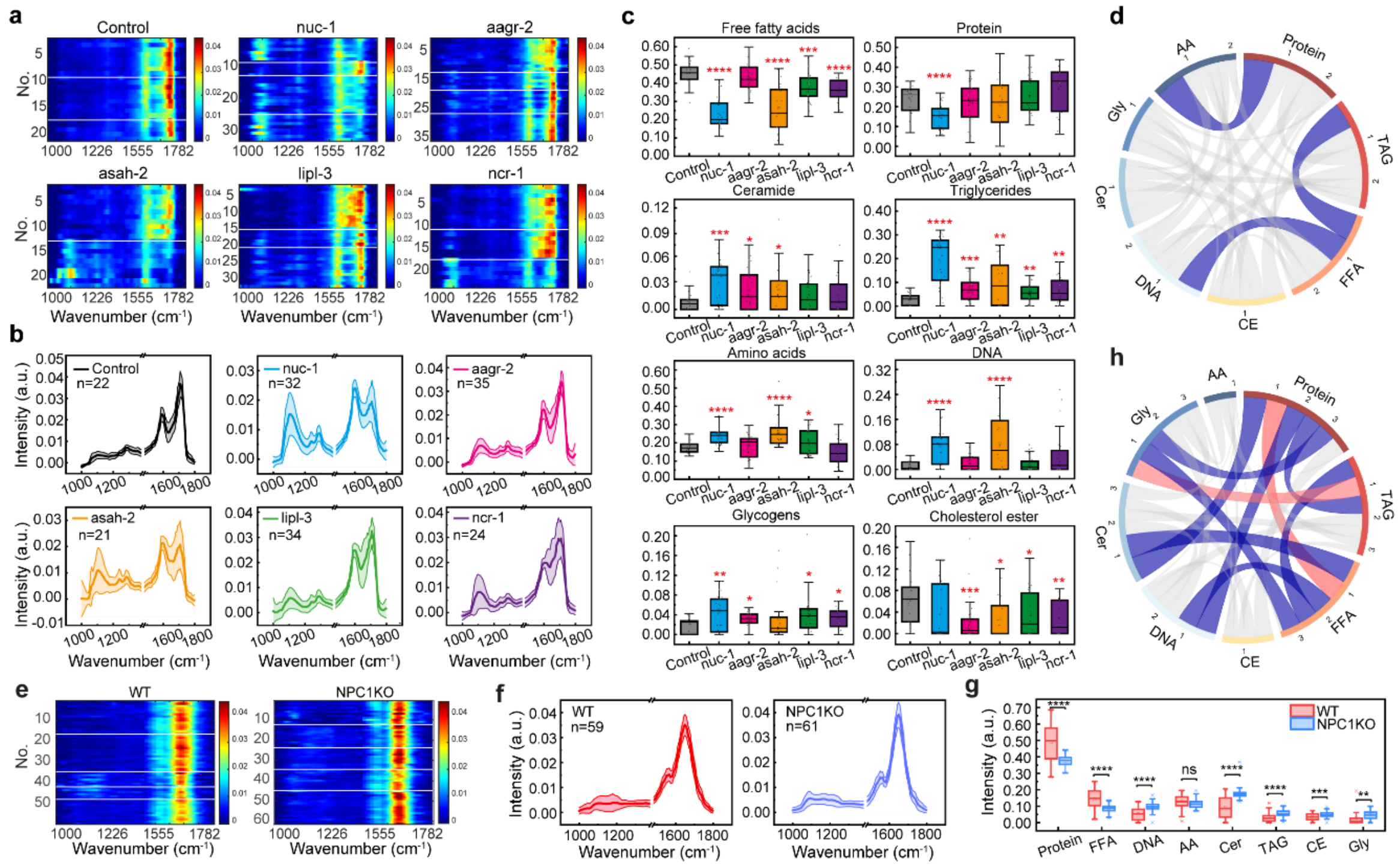

**Figure 5 Profiling of metabolic changes associated with lysosomal storage diseases**. **a**, Heatmap of fingerprint spectra extracted from lysosomes under different RNAi conditions (n=22 for control, n=32 for *nuc-1*, n=35 for *aagr-2*, n=21 for *asah-2*, n=34 for *lipl-3* and n=24 for *ncr-1* derived from two to four independent experiments), illustrating spectral variations across groups. Each row represented a lysosomal spectrum. **b**, Representative average spectrum for each RNAi condition, highlighting distinct metabolic profiles associated with specific RNAi treatments. Shaded area indicates the standard deviation. **c**, High-content analysis of lysosomal contents across RNAi groups, revealing differences in chemical composition and metabolic activity. All comparisons were made relative to the Day 2 control group, whose lysosomal profiles represent the normal metabolic state. (Two-sample t-test: *, p<0.05; **, p<0.01; ***, p<0.001; ****, p<0.0001; ns, not significance). **d**, Pearson correlation analysis of eight lysosomal contents from *C. elegans* samples visualized using a chord diagram. Blue curves represent negative correlations lower than -0.5, and red curves represent positive correlations higher than 0.5, with curve thickness indicating the strength of the correlation. **e**, Heatmap of fingerprint spectra extracted from WT and NPC1KO of HEK293T cells (n=59 for WT, n=61 for NPC1KO derived from five independent experiments). **f**, Representative average spectra of WT and NPC1KO cell lines. Shaded area indicates the standard deviation. **g**, High-content analysis with statistical comparison of lysosomal chemical contents between WT and NPC1KO groups (Two-sample t-test: *, p<0.05; **, p<0.01; ***, p<0.001; ****, p<0.0001; ns,

not significance). **h**, Pearson correlation analysis of eight lysosomal contents from mammalian cells visualized using a chord diagram. Blue curves indicate negative correlations lower than -0.5, and red curves indicate positive correlations higher than 0.5, with curve thickness reflecting correlation strength. In **c** and **g**, the boxes show the interquartile range (IQR), the centerlines indicate medians and the lines outside the boxes extend to 1.5 times the IQR.

# Methods

## FILM hyperspectral imaging.

The pulsed mid-infrared pump beam is generated by a wavelength-tunable quantum cascade laser (QCL, Daylight Solutions, MIRcat-QT-Z-2400). Fluorescence excitation light is provided by either a 488 nm fixed-wavelength diode laser module (Cobolt, 06-MLD 488 nm) or a femtosecond laser (Insight DeepSee, Spectral Physics, Insight DS DUAL), depending on the fluorophore used. The 1040 nm output of the femtosecond laser is frequency-doubled using an LBO crystal and temporally broadened with SF57 rods to generate picosecond 520 nm light. The 488 nm laser can be digitally modulated into pulsed light via an external trigger, while the 520 nm laser is modulated using an acousto-optic modulator (AOM). A function generator synchronizes the visible excitation light and the mid-IR pump beam, with their modulation frequencies set to $2f$ (400 kHz) and $f$ (200 kHz), respectively. The IR pulse width is set to 200 ns, and the visible light operates with a 30% duty cycle. The fluorescence excitation light is rapidly scanned using a pair of dual-axis galvo mirrors (GVS002, Thorlabs). After passing through a scan lens (f = 100 mm; a pair of AC508-100-A, Thorlabs) and a tube lens (f = 200 mm; TTL200-A, Thorlabs), the beam is reflected by a dichroic mirror (DM) into a water-immersion objective (UPlanSApo, Olympus, 60×, NA=1.2) and focused onto the sample. The IR beam is scanned independently with another pair of X-Y galvanometer mirrors (GVS002, Thorlabs). The IR beam path employs a concave mirror as the scan lens (f = 200 mm; CM508-200-P01, Thorlabs) and a tube lens (f = 500 mm; CM508-500-P01, Thorlabs) to relay the scan to the back pupil of a reflective objective (PIKE, 40×, NA=0.78), achieving counter-propagation alignment with the visible excitation light. Before imaging, the IR beam is carefully aligned to overlap with the visible focus. During imaging, the IR and visible foci are synchronously scanned, ensuring uniform excitation and detection over the FOV. The two galvanometer pairs are synchronized with the focal lengths of the visible and IR objectives and scaled based on the beam expansion ratio of the relay system. This scaling factor is calibrated at the start of the experiment. The backward fluorescence emitted from the sample is collected by the water-immersion objective and directed through the DM. After further filtering with a bandpass or long-pass filter, the fluorescence signal is detected by a silicon photomultiplier (SiPM, Hamamatsu, C13366-3050GA). The resulting electrical signal is fed into Moku:Pro (Liquid Instrument, Multi-instrument Mode), filtered, and input into the slots of two lock-in amplifiers for demodulation at $2f$ and $f$ frequencies, corresponding to the FILM and fluorescence DC signals, respectively. These demodulated signals are simultaneously acquired through two input ports of an acquisition card, enabling real-time dual-channel imaging. To perform hyperspectral imaging, the quantum cascade laser (QCL) operates in Multi-Spectral mode using a preset scanning list that that spans the entire fingerprint region. For *S. aureus* imaging, the hyperspectral range covers 980 to 1780 $cm^{-1}$ with 160 frames, while for organelle imaging, including lysosomes, lipid droplets, and mitochondria, it covers 1000 to 1800 $cm^{-1}$ with 126 frames.

**IR spectral calibration.**

Since the signal of FILM is proportional to the DC fluorescence intensity, IR light power and IR absorption cross section of the molecules as described below:

$$Signal = Fluorescence_{DC} * I_{IR} * \sigma_{IR}$$

To obtain the IR absorption spectrum of the molecules, $\sigma_{IR}$, we need to calibrate the spectrum of the collected FILM.

As depicted in **Supplementary Fig. 12**, the raw FILM spectra were initially corrected for the noise-induced baseline and then divided by the fluorescent photobleaching curve. The baseline was estimated using the average intensity under the IR-off condition and approximated by the wavenumber at the IR power dip. Subsequently, the spectra were divided by the IR power spectrum to calibrate the peak resulting from the power profile. Finally, the spectra were smoothed with 3-5 pixels neighboring average and normalized with the area under the curve. Since there is a power dip around 1450 $cm^{-1}$ caused by the switching of laser chips, which may introduce artifacts during power calibration, the spectral band from 1380 to 1480 $cm^{-1}$ was excluded from the quantification analysis.

**Self-supervised hyperspectral denoising.**

To suppress the spatially correlated and spectrally varied noise intrinsic to hyperspectral FILM imaging, we implemented a self-supervised denoising framework termed SPEND (Self-Permutation Noise2Noise Denoising)[30]. In contrast to conventional approaches that depend on explicit noise modeling or on high-SNR reference data for training, SPEND can be trained directly from a single noisy hyperspectral stack.

The key of SPEND is to learn the noise characteristics using two independent measurements of the same field of view (FOV). To achieve this, we proposed a permutation-based method accompanied by correlation identification. Because the scanning step size is much smaller than the features of interest both spectrally and spatially, adjacent pixels or frames can be treated as independent measurements of the same FOV. However, chemical imaging often contains correlated noise, which violates the independence assumption. In our case, noise correlation was primarily observed along the spatial domain (**Supplementary Fig. 4**). Comparison along the spectral axis, which is perpendicular to the correlation axis can help to avoid the noise leakage into signals. Accordingly, the raw hyperspectral stack was split into odd and even frames along the spectral axis, recombined into two complementary sequences, and used as

input–target pairs for network training. These paired sequences could thus be regarded as independent noisy measurements of the same underlying signal.

A 3D U-Net architecture was adopted as the backbone network. Its encoder–decoder structure, comprising convolution, max-pooling, and upsampling layers, efficiently captures both spatial and spectral features even with limited training data. During training, the network learns to map one noisy sequence onto its paired counterpart, thereby suppressing noise while retaining consistent structural and spectral information.

In the prediction phase, the original (non-permuted) hyperspectral stack is fed into the trained model, thereby preserving both spectral continuity and spatial integrity. This approach enables effective removal of spatially correlated and spectrally heterogeneous noise, yielding denoised hyperspectral datasets with substantially improved signal-to-noise ratio. The enhanced data quality facilitates downstream analyses, including ratiometric mapping, spectral unmixing, and quantitative profiling of biochemical contents.

As a self-supervised method based on a permutation strategy, SPEND does not require external labels during training. The trained model can then be applied to new hyperspectral datasets with similar SNR, including those involving different dyes or biological samples, without the need to modify the network architecture. In our implementation, the training set consists of six fields of view, each with dimensions of 200 × 200 × 126 pixels. The training process takes ~30 minutes on an NVIDIA RTX 4090 24GB GPU. Once trained, SPEND performs inference on new hyperspectral image stacks efficiently, requiring ~2 seconds per stack.

**Construction of an eight standard reference library for spectral unmixing.**

Lysosomes are widely recognized as the primary catabolic centers of the cell, responsible for degrading various macromolecules. Accordingly, these macromolecules and their building blocks represent the major chemical constituents of lysosomes. To construct a comprehensive and biologically relevant reference spectral library, the standards were selected based on the four major classes of biomolecules: proteins, lipids, carbohydrates, and nucleic acids. For proteins and lipids, we included both the macromolecules and their respective monomers (protein and amino acids, triglycerides and free fatty acids). In contrast, for carbohydrates and nucleic acids, we didn't include the monomers of glycogen and DNA. This decision was based on the fact that, unlike proteins and lipids, their monomers (glucose and nucleotides) exhibit minimal spectral differences from the polymers, lacking clear peak shifts that could serve as reliable diagnostic markers. Moreover, glycogen and DNA better represent the biological context of lysosomal dysfunction, which primarily involves defective macromolecule degradation, as demonstrated in RNAi studies targeting NUC-1 and AAGR-2. Finally, ceramide and cholesterol esters were included because they are well-known metabolites that frequently accumulate in lysosomal storage disorders, making them essential for accurately modeling pathological lysosomal metabolism.

Detailed information for the eight compounds, including their chemical identities and commercial suppliers, is provided in **Supplementary Table 1**. To obtain FILM spectra of the non-fluorescent standards, we used 50 μM rhodamine 6G (R6G) as a fluorescence reporter by dissolving powdered compounds or mixing liquid samples with R6G, followed by measurement on air-dried gels for powdered compounds or on solutions for liquid samples. The characteristic wavenumbers for each standard, together with the corresponding vibrational mode assignments, are listed in **Supplementary Table 2**.

**MCR-LASSO spectral unmixing.**

To enhance unmixing performance, we input the spectrum of the original standard as the initial estimate into the MCR. Additionally, to mitigate the over-adjustment of the reference spectrum by MCR, we adopted an augmentation MCR strategy, incorporating the reference spectrum into the lysosome spectral dataset as an additional constraint. Following this, the corrected spectrum output by the augmented MCR was used as input for the LASSO spectral unmixing. We segmented individual organelles (e.g., lysosomes) and extracted their fingerprinting spectra. Following photobleaching correction and IR power calibration, the calibrated spectra were reshaped into a 2D matrix, with rows representing individual spectra, for LASSO-based concentration decomposition[32].

***C. elegans* strains.**

*C. elegans* N2 strain was obtained from *Caenorhabditis* Genetics Center (CGC). *C. elegans*

strains were maintained at 20˚C on standard NGM agar plates seeded with OP50 *E.coli* (HT115 *E. coli* for RNAi experiments) using standard protocols[53].

***C. elegans* RNAi treatments.**

RNAi clones used in this study were sourced from the RNAi library generated by Dr. Marc Vidal's lab[54], including *aagr-2, asah-2, ncr-1, lipl-3,* and *nuc-1*. Specifically, *nuc-1* encodes acid deoxyribonuclease, and its deficiency contributes to Autoinflammatory-Pancytopenia Syndrome; *aagr-2* encodes acid alpha-glucosidase, whose loss causes Pompe Disease; *asah-2* encodes acid ceramidase, and its defect is associated with Farber Disease; *lipl-3* encodes lysosomal acid lipase, which is involved in Wolman Disease; and *ncr-1* encodes lysosomal cholesterol transporter and its deficiency results in Niemann-Pick Type C Disease.

All RNAi colonies were selected for resistance to both 50 μg ml−1 carbenicillin and 50 μg ml−1 tetracycline, and verified by Sanger sequencing. RNAi bacteria were cultured for 14 hours in LB with 25 μg/ml carbenicillin, then seeded onto RNAi agar plates containing 1 mM IPTG and 50 μg/ml carbenicillin. Each RNAi bacteria clone was allowed to dry on the plates before overnight incubation at room temperature to induce dsRNA expression. RNAi-based experiments were conducted using *E. coli* HT115 bacteria, with L4440 empty vector bacteria used as controls. For RNAi plates containing LysoSensor, LysoSensor was added to RNAi plates at 0.5 μM final concentration.

Synchronized L1 N2 worms were added onto 6 cm RNAi plates and raised at 20 °C for two days till the L4 stage. Around 50 L4 worms were transferred to the RNAi plates containing LysoSensor. At Day 2, worms were anesthetized with 1% $NaN_3$ and imaged using FILM, mounted between a glass coverslip and a $CaF_2$ substrate.

***C. elegans* aging experiments.**

Synchronized L1 N2 worms were added onto 6 cm plates and raised at 20 °C. Around 50 worms at L4, day 2, day 4, day 8 were transferred to LysoSensor containing plates and imaged at day 2, 4, 6 and 10, respectively.

**Physiological validation of the 1587 and 1711 cm$^{-1}$ spectral features.**

A functional perturbation experiment by knocking down the lysosomal cystine transporter cystinosin (CTNS-1)[35], whose loss is known to cause cystine accumulation in the lysosome was performed to further validate the 1587 $cm^{-1}$ assignment of amino acids. As shown in **Extended Data Fig. 5a**, FILM imaging revealed a modest but statistically notable increase in the lysosomal amino acid signal following CTNS-1 knockdown ($P = 0.013$, two-sided two-sample t-tests). This effect likely reflects the selective impact of CTNS-1 loss on cystine rather than a global change in all amino acids. This physiological validation provides independent evidence supporting our interpretation of this spectral feature as a readout of lysosomal amino acid levels.

A transgenic C. elegans strain with upregulated lysosomal lipase LIPL-4 (*lipl-4 Tg*), which is known to enhance lysosomal lipolysis, based on lysosome-targeted lipidomics using mass spectrometry (MS)[36], was used to validate the ratiometric measurement of lipolytic activity and the 1711 $cm^{-1}$ assignment of fatty acids. As shown in the **Extended Data Fig. 5b-c**, FILM imaging reveals higher levels of fatty acids ($P = 2.14\times10^{-52}$, two-sided two-sample t-tests) and increased lipolytic activity ($P = 3.11\times10^{-18}$, two-sided two-sample t-tests) in the lysosomes of *lipl-4 Tg* worms compared to wild type controls. The consistency with the MS-based lysosome-targeted lipidomics results supports the validity of the FILM approach for *in vivo* assessment of lysosomal metabolic activity.

**Cell line.**

HeLa cells were purchased from the American Type Culture Collection (ATCC). HEK293T sgNT (NPC1 +/+, control) and sgNPC1 (NPC1 -/-, knockout) cells were a gift from Dr. Roberto Zoncu (University of California, Berkeley; PMID: 33308480[55]). Mycoplasma contamination was regularly tested, and cells were confirmed to be mycoplasma-free using MycoAlert Mycoplasma Detection Kit (Lonza, LT07-318).

**Bacteria strains.**

*Shigella flexneri* expressing GFP was grown overnight at 37 °C on a tryptic soy agar plate. Colonies with green fluorescence were picked up by sterile inoculation loops and then resuspended in PBS. The bacterial solution was diluted by optical density at 600 nm (OD600) to 0.1. The bacteria were then fixed by 10% formalin for 30 minutes at room temperature. The bacterial solution was washed twice with deionized water, dried on $CaF_2$, and sandwiched with a coverslip. The samples were then observed using the FILM setup with a DM (500 LP, Edmund, #69-899) and a filter (520/36, Edmund, #67-016).

*Staphylococcus aureus* (*S. aureus*) was incubated in a MHB medium for 10 h. After centrifuging and washing in phosphate-buffered saline (PBS), the bacteria were fixed by formalin solution for 30 minutes. Rhodamine 6G at $10^{-4}$ M was then added to the bacteria pellet, which was subsequently resuspended and incubated for 1 hour. Following the final washing steps with deionized water, the bacterial suspension was dried on $CaF_2$ and sandwiched with a coverslip for FILM imaging with a DM (550 LP, Edmund, #69-900) and filter (575/27, Edmund, #33-333).

**Spectral phasor analysis.**

In spectral phasor analysis, the spectrum of each pixel was interpreted through the discrete Fourier transform of first-order harmonics. By scattering the pixels of the entire image across the complex plane, we were able to identify specific clusters that represented target chemical channels. Phasor analysis was performed with the standardized phasor analysis plug-in in ImageJ (1.49v). The phasor domain segmentation was shown in **Supplementary Fig. 9**.

**Fluorescence labeling of mammalian cells.**

*HeLa cells labelled by Lipi-Red*: HeLa cells purchased from the American Type Culture Collection (ATCC) were seeded on $CaF_2$ substrates at a density of $1 \times 10^5$ cells/ml in 2 ml of high-glucose DMEM supplemented with 10% FBS and penicillin–streptomycin and incubated for 24 h at 37°C in a humidified atmosphere with 5% $CO_2$. The following day, the medium was replaced with fresh serum-free medium containing 6 μM Lipi-Red (LD03, DOJINDO), and the cells were incubated at 37°C for 30 minutes. After incubation, the cells were gently washed three times with warm PBS to remove excess dye. For imaging, the cells on $CaF_2$ are sandwiched with coverslip, maintained in PBS and observed using FILM setup with DM (550 LP, Edmund, #69-900) and filter (600 LP, Edmund, #62-985).

*HeLa cells labelled by MitoTracker™ Green*: HeLa cells purchased from the American Type Culture Collection (ATCC) were seeded on $CaF_2$ substrates at a density of $1 \times 10^5$ cells/ml in 2 ml of high-glucose DMEM supplemented with 10% FBS and penicillin–streptomycin and incubated for 24 h at 37°C in a humidified atmosphere with 5% $CO_2$. The following day, the medium was replaced with fresh serum-free medium containing 100 nM MitoTracker Green (M7514, Thermo Fisher Scientific), and the cells were incubated at 37°C for 15 minutes. After incubation, the cells were gently washed three times with warm PBS to remove excess dye. For imaging, the cells on $CaF_2$ are sandwiched with coverslip, maintained in PBS and observed using FILM setup with DM (500 LP, Edmund, #69-899) and filter (520/36, Edmund, #67-016).

*HEK293T cells labelled by LysoSensor$^{TM}$ DND189*: WT and NPC1KO HEK293 cells were seeded on $CaF_2$ substrates at a density of $1 \times 10^5$ cells/ml in 2 ml of high-glucose DMEM supplemented with 10% FBS and penicillin–streptomycin and incubated for overnight at 37°C in a humidified atmosphere with 5% $CO_2$. The following day, the medium was replaced with medium containing 1 μM LysoSensor DND189 (L7535, Thermo Fisher Scientific), and the cells were incubated at 37°C for 20 minutes. After incubation, the cells were gently washed three times with warm PBS to remove excess dye. For imaging, the cells on $CaF_2$ are sandwiched with coverslip, maintaining in PBS and observed using FILM setup with DM (500 LP, Edmund, #69-899) and filter (520/36, Edmund, #67-016).

**ATR-FTIR spectroscopy.**

The FTIR spectra of all samples were measured on an attenuated total reflection (ATR) FTIR spectrometer (Nicolet Nexus 670, Thermo Fisher Scientific). The measurements were conducted with a spectral resolution of 2 $cm^{-1}$, and each spectrum was measured with 32 scans. Prior to measurement, the ATR crystal was carefully cleaned with ethanol and dried to prevent contamination between samples. All spectra were automatically normalized using the built-in baseline correction feature of the spectrometer.

**Thermal sensitivity measurement of fluorescence dye.**

The thermal sensitivity measurement of fluorescent dye was measured with the setup shown in **Supplementary Fig. 3a**. A small droplet of fluorescent dye solution was placed on a silicon wafer, which served as the substrate. A coverslip was then positioned on top to ensure the sample was securely enclosed and evenly distributed. The sample is excited by light from LED (SOLIS-3C, Thorlabs), which passes through an excitation filter to select the specific wavelength required to excite the dye. The excitation light was directed toward the sample through the optical setup. The emitted fluorescence was collected and passed through an emission filter, which blocks any residual excitation light and isolates the specific fluorescence wavelength of interest. This filtered emission light was then detected by camera (Hamamatsu, C13440). Since the sample was placed on a temperature-controlled plate (Bioscience Tools, TC-1-100s), we can control the temperature and analyze the corresponding changes in fluorescence intensity.

**Statistics analysis.**

In the aging studies, we analyzed lysosomal function over time in live, anesthetized wild-type worms at four different ages (Day 2, Day 4, Day 6, and Day 10). This analysis revealed age-dependent metabolic changes at the lysosomal level. Approximately six worms were imaged per age group, and ~10 lysosomes per intestinal cell were analyzed to provide sufficient statistical capability. For disease-associated studies, all comparisons were made relative to carefully matched wild-type controls. Control data are shown in all panels, labeled as “Control” in panels a, b, and c, and as “WT” in panels e, f, and g. For each condition, three to five independent biological replicates were performed, with 5-10 lysosomes analyzed per replicate, ensuring statistical robustness of the comparisons.

All significance analyses were performed using two-sided two-sample t-tests. *, $p<0.05$; **, $p<0.01$; ***, $p<0.001$; ****, $p<0.0001$; ns, not significance.

## Data availability

All data are available in the main paper or supplementary materials have been deposited via figshare at https://doi.org/10.6084/m9.figshare.31302607 (Ref.[56]).

## Code availability

Code supporting the current study, including SPEND denoising, and spectral unmixing, can be accessed at the Cheng Lab GitHub page (https://github.com/buchenglab) under the GNU General Public License v3.0. The SPEND is also available via Zenodo at https://doi.org/10.5281/zenodo.19236367 (ref.[57]).

## Extended Data Figures:

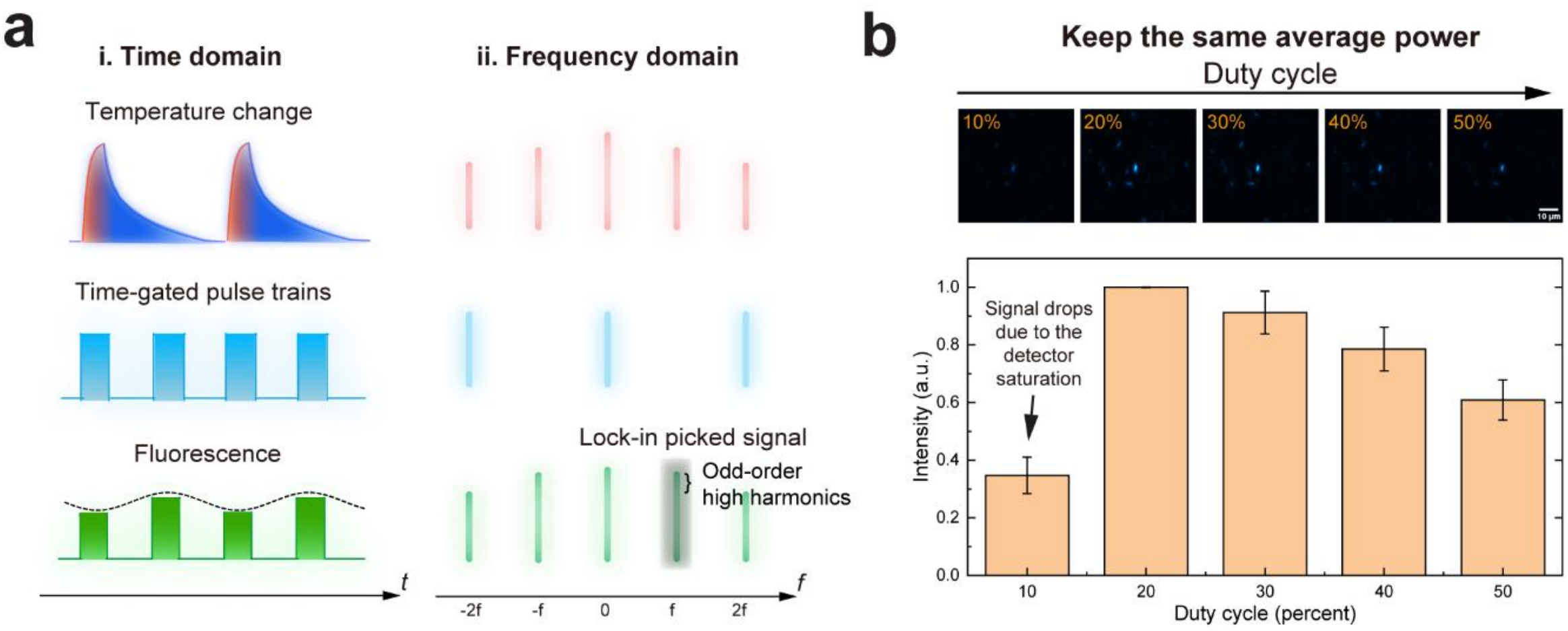

**Extended Data Fig. 1: Optical boxcar strategy enhances the signal by shifting high odd-order harmonics to detected frequency.**

**a,** Principle of the higher order harmonics shifting. The fluorescence excitation light was modulated into pulses. The pulsed excitation light functioned as a 2f carrier, shifting high odd-order harmonic signals into the demodulation frequency. **b,** FILM signal of *Shigella flexneri* expressing GFP at different duty cycle (n=5). Statistical data are presented as mean ± s.d. Scale bar: 10 μm.

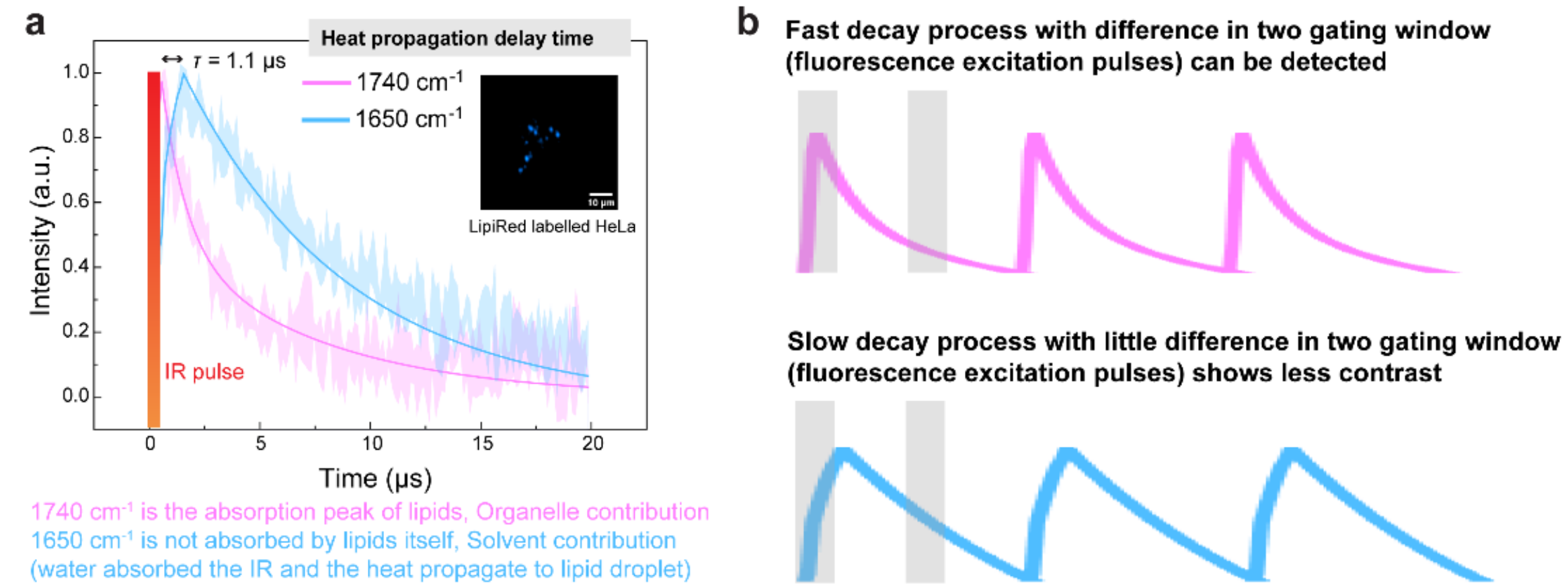

**Extended Data Fig. 2: Optical boxcar strategy suppressed solvent background by harnessing the differential thermal dynamics between particles and water medium.**

**a,** Photothermal dynamics of LipiRed labelled lipid droplets under different IR absorption peaks. The shaded area represents the mean derived from three independent measurements, while the solid line indicates the exponential fitting curve. **b,** The pulse pair serves as the gating windows to capture the time-resolved fluorescence signal, which is less sensitive to slow dynamic processes. Scale bar: 10 μm.

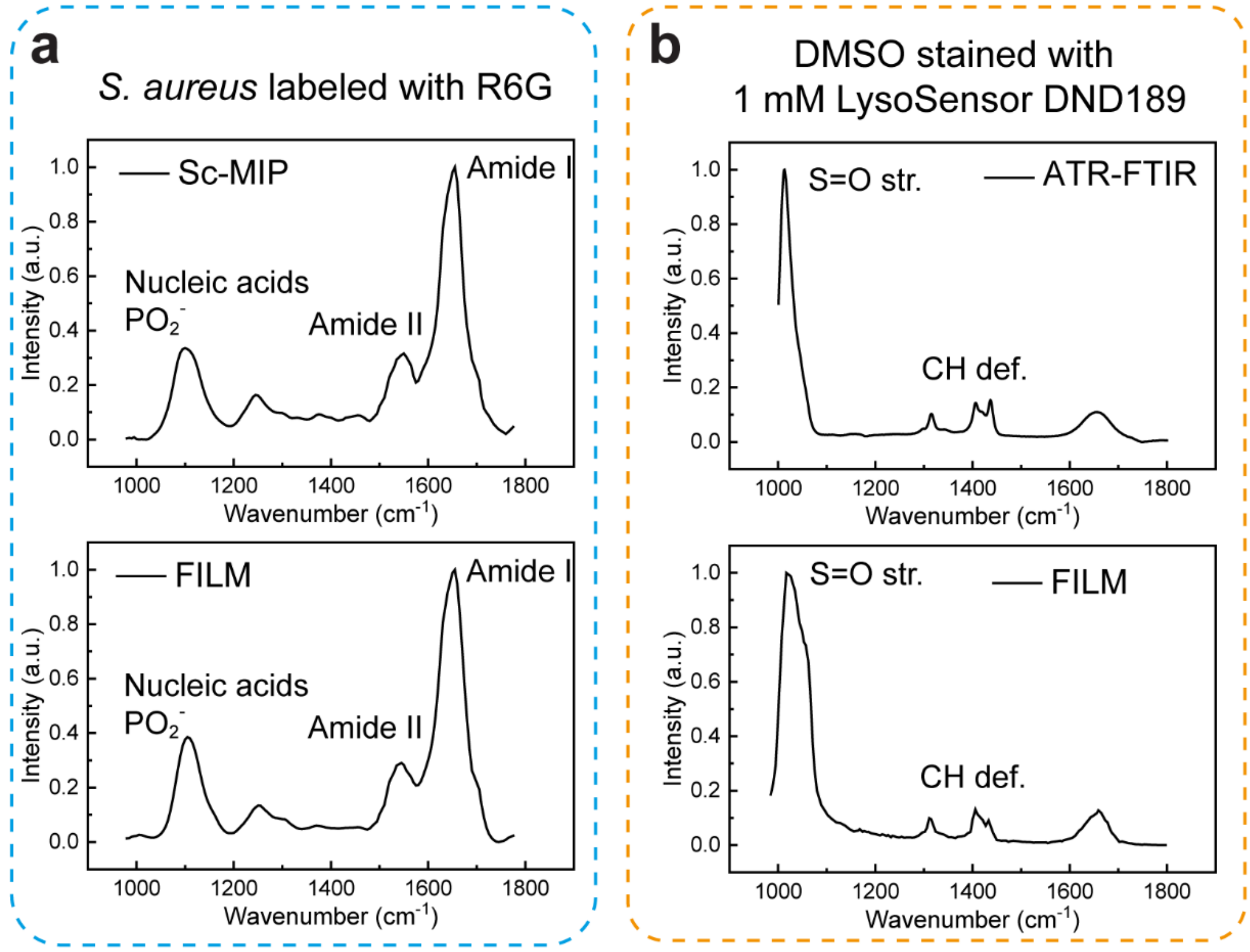

**Extended Data Fig. 3: Spectral fidelity verification.**

**a,** Spectral comparison of FILM and scattering-based MIP (Sc-MIP) with Rhodamine 6G labelled *S. aureus*. **b,** Spectral comparison of FILM and ATR-FTIR with LysoSensor DND189 stained DMSO.

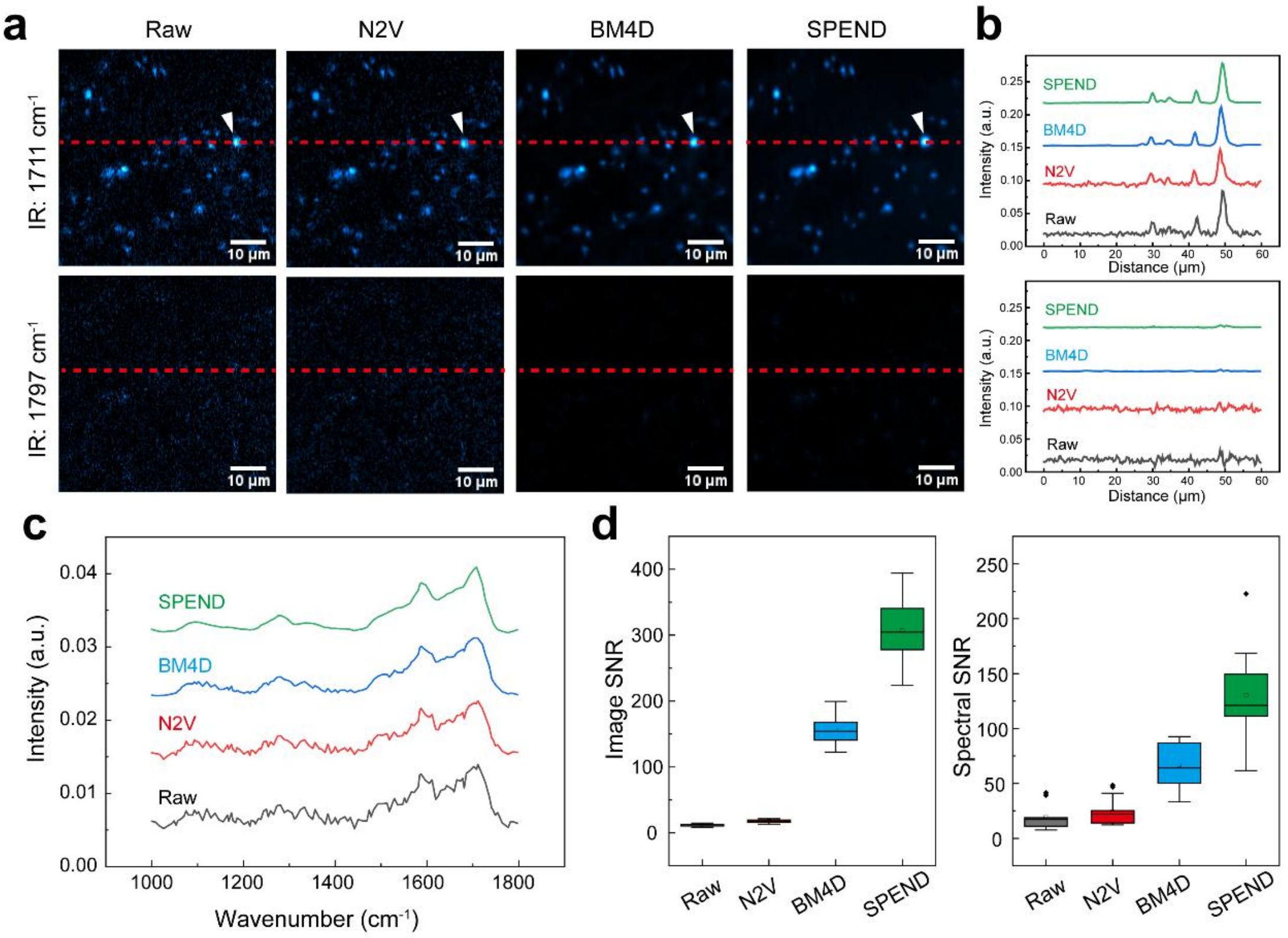

**Extended Data Fig. 4: Head-to-head comparison of SPEND with BM4D and Noise2Void (N2V).**

**a**, FILM images of lysosomes were acquired with IR at 1711 cm⁻¹ and 1797 cm⁻¹ before and after three denoising algorithms. **b,** the intensity profiles along the red dotted lines indicated in **a**. **c,** compares the raw, uncalibrated FILM spectra before and after three denoising algorithms. **d,** the quantification of image SNR and spectral SNR before and after three denoising algorithms (n = 13). In **d**, the boxes show the IQR, the centerlines indicate medians and the lines outside the boxes extend to 1.5 times the IQR. Scale bar: 10 μm.

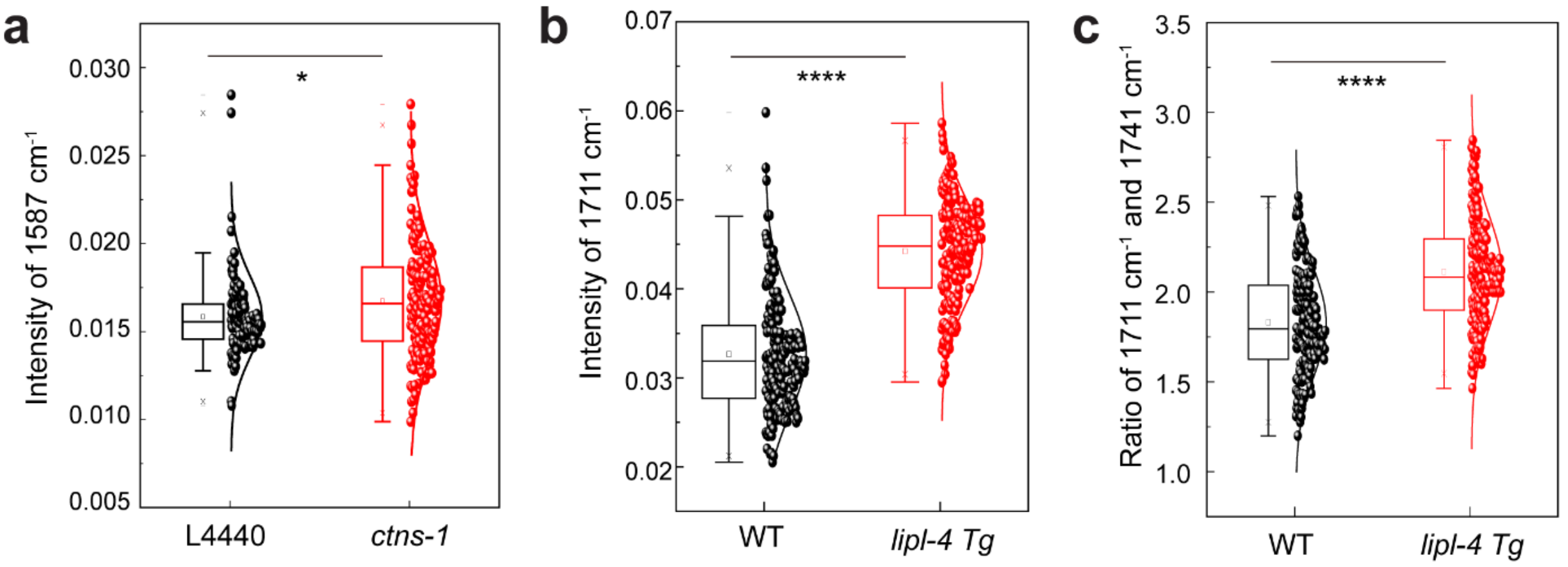

**Extended Data Fig. 5: Physiological validation of the 1587 and 1711 cm$^{-1}$ features.**

**a,** Quantification of the 1587 cm$^{-1}$ intensity for individual lysosomes. CTNS-1 knockdown strain showed a modest but statistically significant increase in the lysosomal amino acid signal. A significant increase in signal intensity at 1587 cm$^{-1}$ is observed in *ctns-1* compared to L4440 worms (Two-sample t-test: *, p=0.013). **b,** Quantification of the 1711 cm$^{-1}$ peak intensity shows a significant increase in *lipl-4 Tg* worms compared to WT. **c,** Ratio of the 1711 cm$^{-1}$ (free fatty acids) to 1741 cm$^{-1}$ (lipid ester) peaks, indicating a relative upregulated in lipolytic activity upon LIPL-4 overexpression. ****p < 0.0001 by two-sample t-test. Each point represents a single lysosome, with box plots indicating the median, interquartile range, and whiskers representing 1.5× IQR.

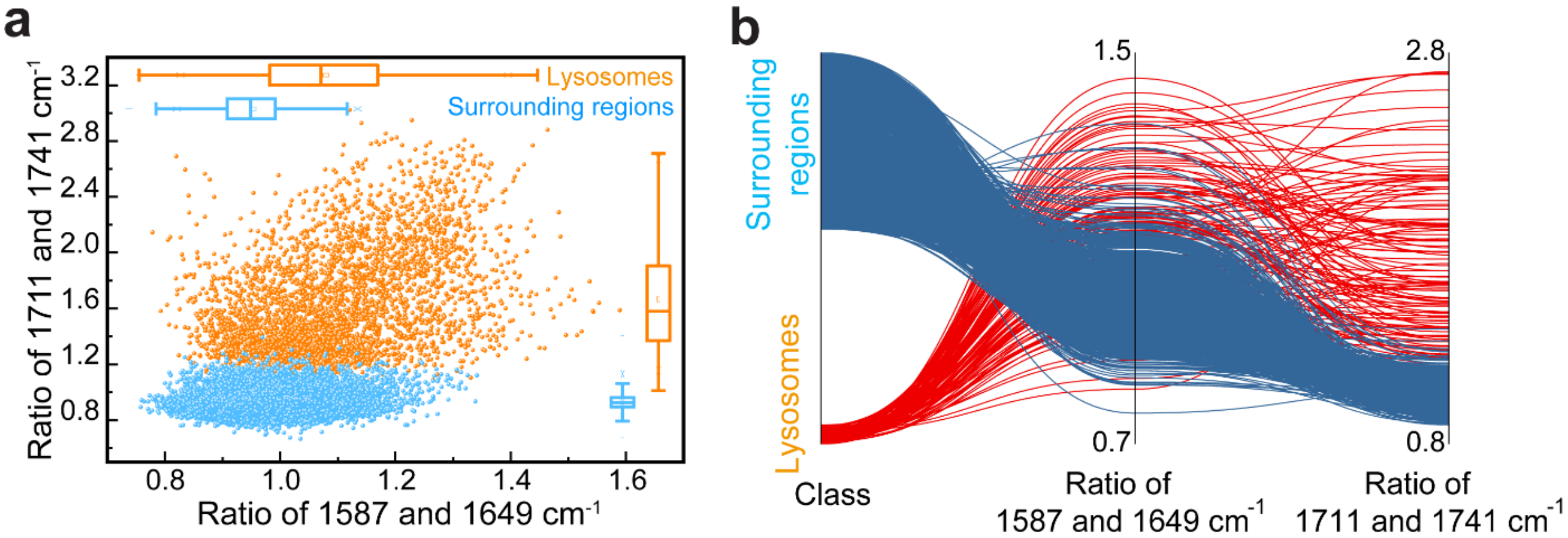

**Extended Data Fig. 6: Visualization of two ratios and class discrimination.**

**a,** Pixel-wise scatter plot of two calculated intensity ratios (200×200 pixels). **b,** The parallel set shows the relationship between two spectral ratios (1587 cm$^{-1}$/1649 cm$^{-1}$ and 1711 cm$^{-1}$/1741 cm$^{-1}$) and the class separation of lysosomes (red) and surrounding region (blue). The curves represent different data points from the corresponding classes, illustrating the distribution and class distinction.

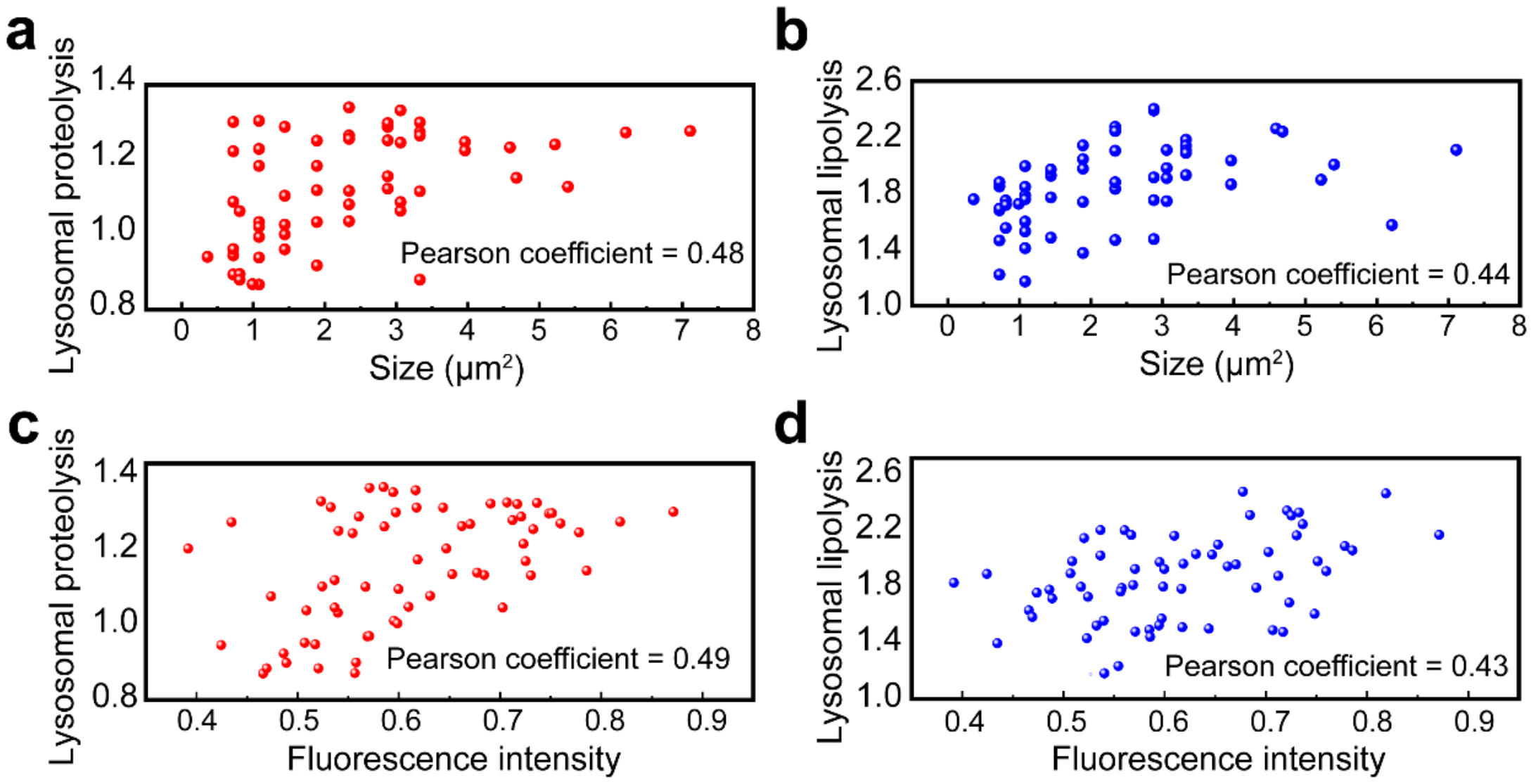

**Extended Data Fig. 7: Correlations of lysosomal hydrolytic activity with lysosomal size and fluorescence intensity.**

**a,** The correlation between lysosomal proteolytic activity and size is not significant with a Pearson coefficient of 0.48. **b,** The Pearson coefficient of lysosomal lipolytic activity and size is 0.44. **c,** The correlation between lysosomal proteolytic activity and fluorescence intensity is not significant with a Pearson coefficient of 0.49. **d,** The Pearson coefficient of lysosomal lipolytic activity and fluorescence intensity is 0.43.

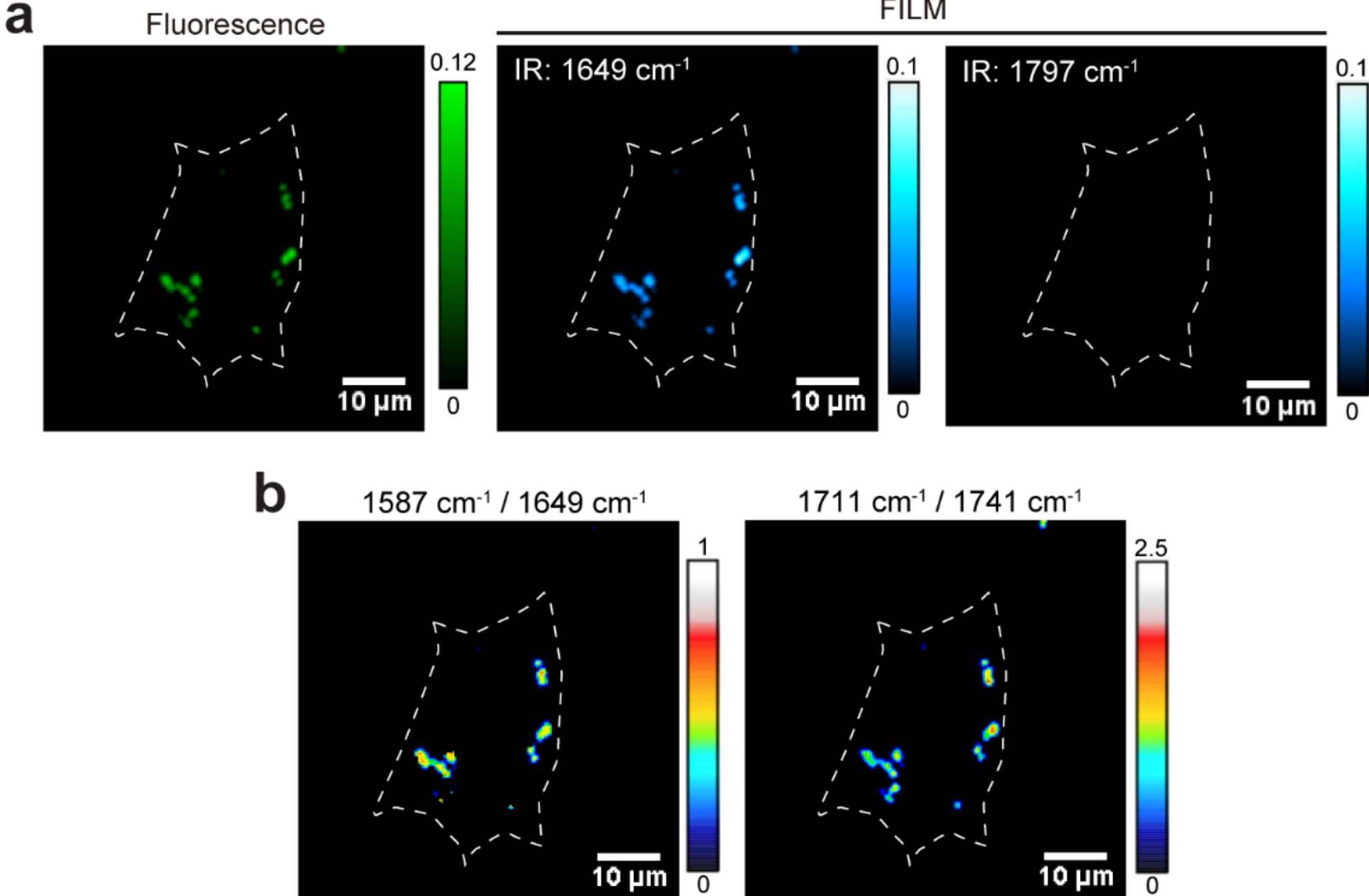

**Extended Data Fig. 8: Hydrolytic heterogeneity of lysosomes in mammalian cells.**

**a,** Fluorescent and FILM images at 1587 and 1649 $cm^{-1}$ of LysoSensor DND189 labelled HEK293T cells. **b,** Ratio-metric mapping of intensity ratios at 1587/1649 $cm^{-1}$ and 1711/1741 $cm^{-1}$. Scale bar: 10 μm. Representative results are shown from three independent experiments.

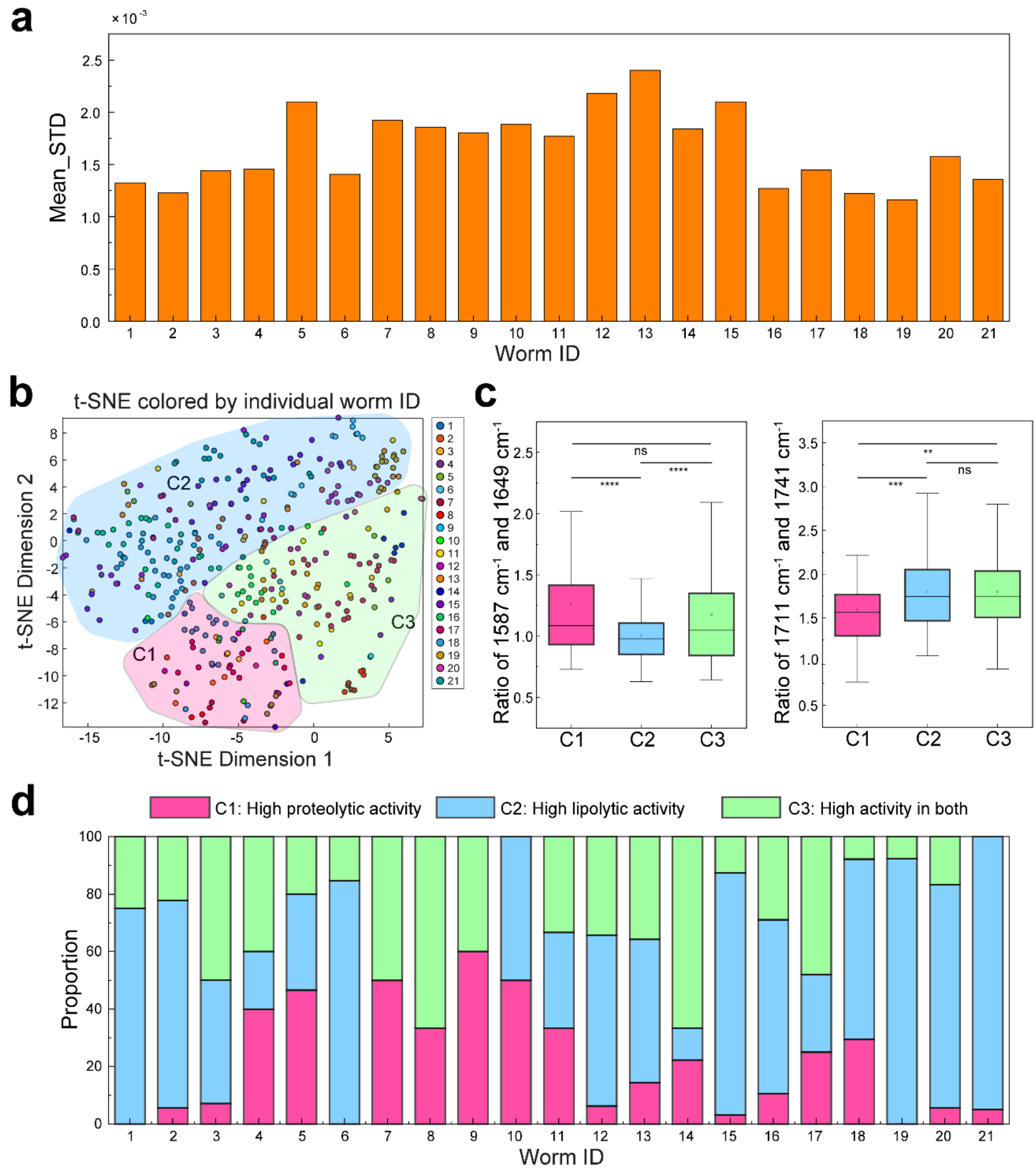

**Extended Data Fig. 9: t-SNE–based classification of WT lysosomal spectra and identification of functional subtypes.**

**a,** Within-individual heterogeneity. For each spectrum, the standard deviation (STD) across wavenumbers was calculated; Mean_STD denotes the average STD per worm (y-axis $\times10^{-3}$, x-axis = Worm ID). **b,** t-SNE embedding (one dot per spectrum) colored by individual worm. Three regions are drawn: C1 (red), C2 (blue), C3 (green), corresponding to three clusters of "high proteolytic activity", "high lipolytic activity", and "high activity in both". **c,** Cluster-specific quantitative analysis of activity ratios at 1587/1649 $cm^{-1}$ (proteolytic activity) and 1711/1741 $cm^{-1}$ (lipolytic activity), confirming functional differences among clusters; colors match panel **b**. **d,** Per-worm composition across regions. For each worm, the fraction of its spectra falling into C1/C2/C3 is shown as a stacked bar; colors match panels **b**-**c**. Two-sample t-test: *, $p<0.05$; **, $p<0.01$; ***, $p<0.001$; ****, $p<0.0001$; ns, not significance. In **c**, the boxes show the IQR, the centerlines indicate medians and the lines outside the boxes extend to 1.5 times the IQR.

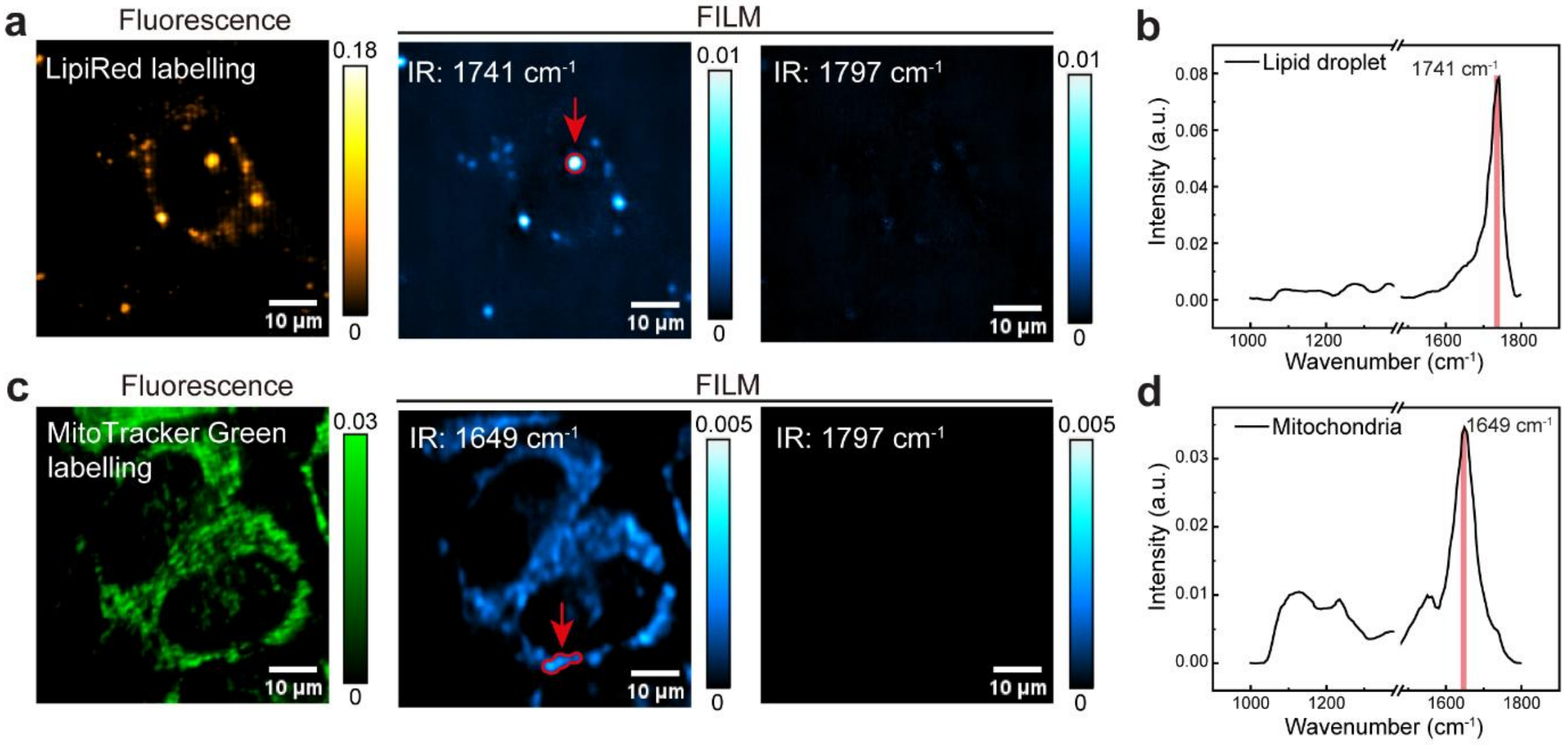

**Extended Data Fig. 10: Hyperspectral FILM imaging of lipid droplet and mitochondria.** **a,** Fluorescence and FILM images at 1741 (ester C=O, on-resonance) and 1797 $cm^{-1}$ (off-resonance) of LipiRed labelled HeLa cells. **b,** FILM spectral of lipid droplet marked by red circle and arrowhead in **a**. **c,** Fluorescence and FILM images at 1649 (Amide I, on-resonance) and 1797 $cm^{-1}$ (off-resonance) of MitoTracker Green labelled HeLa cells. **d,** FILM spectral of mitochondria marked by red circle and arrowhead in **c**. Scale bar: 10 μm. Representative results are shown from three independent experiments.

# Table of Content

## Supplementary Notes

## Supplementary Figures

## Supplementary Table

## Supplementary Videos

# Supplementary notes:

**Supplementary note 1: Pulsed mode enhanced signal by shifting high odd-order harmonics.**

As depicted in **Extended Data Fig. 1**, mid-infrared photothermal signals generated by periodic IR pulses exhibit a low-duty-cycle characteristic, resulting in a spectrum rich in high harmonic content. This signal can be expressed as a Fourier series:

$$s(t) = \sum_{n=0}^{\infty} A_n e^{j2\pi nft}$$

Where, $nf$ is the $n^{\text{th}}$ harmonic components. $A_n$ is the Fourier coefficient, which reflects the amplitude of each harmonic. A lock-in amplifier (LIA) selectively detects only the first harmonic component.

Pulsed visible light can be described as a square wave, given by:

$$x(t) = \sum_{k=0}^{\infty} B_k \cos\,(2\pi k f_c t)$$

Where, $kf_c$ is the $k^{\text{th}}$ harmonic components. $B_k$ is the Fourier coefficient.

When the photothermal signal is gated by the pulsed visible light, the detected signal is given by:

$$d(t) = s(t) \cdot x(t) = \sum_{n=0}^{\infty} A_n e^{j2\pi nft} \cdot \sum_{k=0}^{\infty} B_k \cos(2\pi k f_c t)$$

$$= \sum_{n=0}^{\infty} \sum_{k=0}^{\infty} A_n B_k e^{j2\pi nft} \cos(2\pi k f_c t)$$

$$= \sum_{n=0}^{\infty} \sum_{k=0}^{\infty} A_n B_k \frac{1}{2} \left(e^{j2\pi(nf + kf_c)t} + e^{j2\pi(nf - kf_c)t}\right)$$

Since the visible light is modulated at twice the frequency of the IR pulses (i.e., $f_c = 2f$), leading to

$$d(t) = \sum_{n=0}^{\infty} \sum_{k=0}^{\infty} A_n B_k \frac{1}{2} \left(e^{j2\pi(n+2k)ft} + e^{j2\pi(n-2k)ft}\right)$$

From this expression, we observe that when $k = 1$, third harmonic component in $s(t)$ (i.e., $n = 3$) is shifted into the first harmonic band since $n - 2k = 1$. Similarly, fifth harmonic is shifted when $k = 2$ and higher harmonics follow the same pattern.

Principally, there is a trade-off: shorter pulses shift more higher harmonics, while the overall decrease in energy reduces the amplitude of each original frequency component (**Supplementary Fig. 2**). As a result, within the "stable zone" the signal reaches its peak when the duty cycle is set to approximately 50% (**Fig. 2f**). If we maintain the average power while reducing the duty cycle, the ideal scenario would be to match the peak temperature. However, due to the low photon budget of silicon photomultiplier (SiPM), the detector easily saturates, disrupting demodulation and ultimately reducing the signal (**Extended Data Fig. 1b**).

**Supplementary note 2: Solvent background suppression with optical boxcar detection.**

We used HeLa cells labeled with LipiRed to mark lipid droplets as a model system. The cells were maintained in phosphate-buffered saline (PBS) solution, and the IR light was set to 1650 $cm^{-1}$ and 1740 $cm^{-1}$, corresponding to the absorption peaks of bulky environmental contribution and lipid droplets (specific to the labeled objects), respectively. The fluorescence dynamics recorded under these conditions are shown in **Extended Data Fig. 2a**. Notably, there is a 1.1 μs delay between the thermal dynamics associated with the surrounding environment (1650 $cm^{-1}$) and those caused by the direct heating of lipid droplets (1740 $cm^{-1}$). Furthermore, the cooling rate of the bulky environment-induced dynamics is significantly slower than that of the lipid droplets. The distinct time delays and prolonged decay characteristics of these two components offer a way to differentiate the actual sample from the surrounding environment. The pulse pair serves as the gating windows to capture the time-resolved fluorescence signal, which is less sensitive to slow dynamic processes, leading to minimal differences between them (**Extended Data Fig. 2b**). This allows for the differentiation and suppression of the surrounding solvent environment's contribution.

**Supplementary note 3: Correlated noise in FILM images.**

FILM employs heterodyne detection, utilizing lock-in amplifier (LIA) to selectively detect and amplify signals at a specific modulation frequency while suppressing noise outside this bandwidth. However, the low-pass filter within the LIA introduces a delayed response, preventing the system from instantaneously tracking rapid variations in the input signal. Consequently, the measured signal at each pixel retains residual influence from its neighboring pixels, leading to temporal and spatial correlation in the detected image noise. Moreover, frequency mismatches between the heterodyne detection system and the reference frequency set in the LIA can further exacerbate this issue. When the detected frequency deviates slightly from the expected reference, the LIA may inadvertently mix signals from adjacent pixels, resulting in unintended signal leakage. This leakage manifests as a structured noise pattern in the image, where neighboring pixels exhibit correlated noise rather than remaining independent. As shown in **Supplementary Fig. 4**. There is an obvious tailing phenomenon between the pixels in the fast-scanning direction of the image, indicating that they are not independent of each other.

**Supplementary note 4: Signal to noise calculation.**

To evaluate the SPEND denoising performance of FILM hyperspectral data (**Fig. 2c-d**), we compared the SNR from both the spatial (image-level) and spectral dimensions.

Image SNR was assessed using the hyperspectral image at 1711 $cm^{-1}$, calculated by the following formula.

$$SNR_{Image} = \frac{\mu_{signal}}{\sigma_{noise}}$$

Where, $\mu_{signal}$ represents the mean pixel intensity of lysosomes. $\sigma_{noise}$ denotes the standard deviation of pixel intensities in a non-lysosome blank area, representing background noise.

Spectral SNR evaluates noise characteristics across the fingerprint for lysosomes. It was assessed in the spectral dimension with the following formula.

$$SNR_{Spectral} = \frac{P_{spectral}}{\sigma_{spectral}}$$

Where, $P_{spectral}$ is the mean intensity of lysosomes at 1711 $cm^{-1}$, which is the strongest peak frame. $\sigma_{spectral}$ represents the standard deviation of intensity values within the same area at 1764, 1777, 1782, 1788, and 1797 $cm^{-1}$, serving as a measure of spectral noise.

**Supplementary note 5: Cosine similarity and Euclidean distance calculation.**

To quantitatively assess the accuracy of spectral unmixing (**Fig. 2f-g**), we employed cosine similarity and Euclidean distance to compare the input spectra with reconstructed results.

Cosine similarity evaluates the angular similarity between two spectral vectors, measuring their directional alignment while ignoring magnitude differences. The cosine similarity between the original input spectrum $S_{orig}$ and the reconstructed spectrum $S_{recon}$ is calculated as:

$$Cosine\ similarity = \frac{S_{orig} \cdot S_{recon}}{\|S_{orig}\| \|S_{recon}\|}$$

Where, $S_{orig} \cdot S_{recon}$ is the dot product of the two spectra and $\|S_{orig}\|$ and $\|S_{recon}\|$ are their respective Euclidean norms. A more accurate LASSO unmixing results in a reconstructed spectrum closer to the original input spectrum, with a cosine similarity approaching 1.

Euclidean distance measures the absolute difference between the original input and reconstructed spectra at each wavenumber, given by:

$$d(S_{orig}, S_{recon}) = \sqrt{\sum_{i=1}^{n} (S_{orig,i} - S_{recon,i})^2}$$

Where, $S_{orig,i}$ and $S_{recon,i}$ are the spectral intensities at the $i^{th}$ wavenumber. $n$ is the total number of spectral data points. A lower Euclidean distance indicates a closer match between the original and reconstructed spectra.

**Supplementary note 6: Lysosomes exhibit distinctive spectral features compared to the surrounding tissues.**

As shown in **Supplementary Fig. 8a**, we observed that at the IR wavenumber of 1587 $cm^{-1}$, the lysosome signal is stronger than at the 1649 $cm^{-1}$; whereas the surrounding background tissue, likely visualized by autofluorescence, exhibited the opposite trend, with higher intensity at 1649 $cm^{-1}$. **Supplementary Fig. 8b** shows the profiles represented by the orange lines in the two images of **Supplementary Fig. 8a**. The difference between these two profiles demonstrates two distinct IR absorption features in space. By segmenting the hyperspectral spectrum with spectral phasor (**Supplementary Fig. 9a-b**), lysosomes could be well identified from the surrounding tissues, and the resulting spectrum is shown in **Supplementary Fig. 9c**. The surrounding tissue exhibited three characteristic peaks at 1553, 1649, and 1741 $cm^{-1}$, corresponding to the amide II, amide I, and ester C=O vibrations of proteins and lipids, respectively[1]. In contrast, the lysosome spectrum concentrated around 1587 $cm^{-1}$ and 1711 $cm^{-1}$.

**Supplementary note 7: Investigation of the peak assignment of lysosomal signatures.**

As shown in **Supplementary Fig. 8c**, we found that, compared to the Amide I peak of protein at ~1649 $cm^{-1}$, the main peak for AA locates at ~1587 $cm^{-1}$, which matches the lysosomal signature. This peak of AA reflects the antisymmetric deformation of $NH_3^+$ and the antisymmetric stretching vibration of $COO^-$ bonds[2]. Lysosomes contain diverse hydrolases that digest macromolecules, breaking proteins into AA and lipid esters into free fatty acids (FFA). Upon opening peptide bonds in proteins, the exposure of additional $NH_3$ and COOH functional groups leads to an IR absorption shift from 1649 $cm^{-1}$ to 1587 $cm^{-1}$. Similarly, the hydrolysis of lipid ester into FFA causes the stretching vibration mode of C=O red-shifting from 1741 $cm^{-1}$ to 1711 $cm^{-1}$.

## Supplementary Figures:

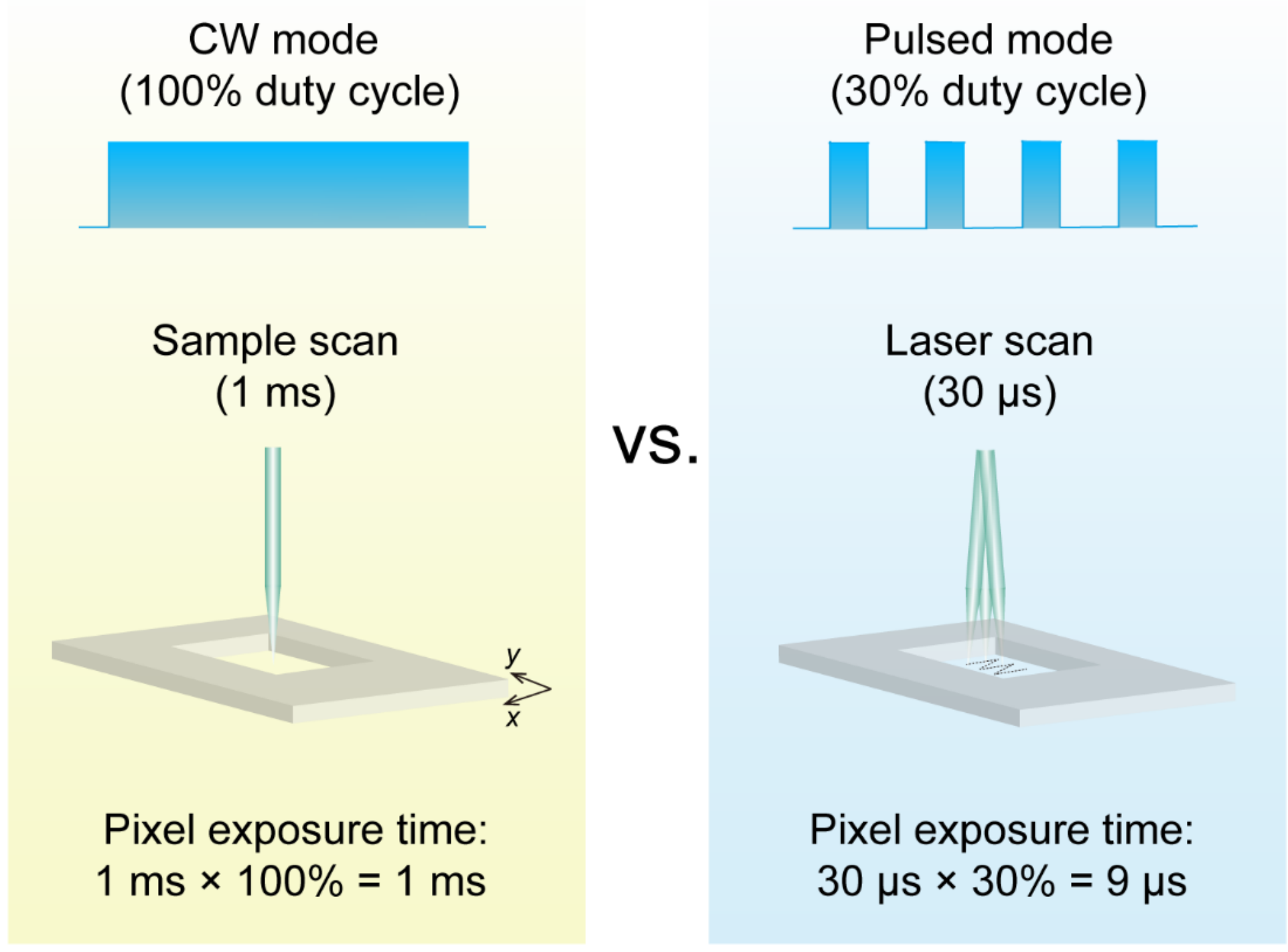

**Fig. S1:** Comparison of fluorescence excitation light exposure durations.

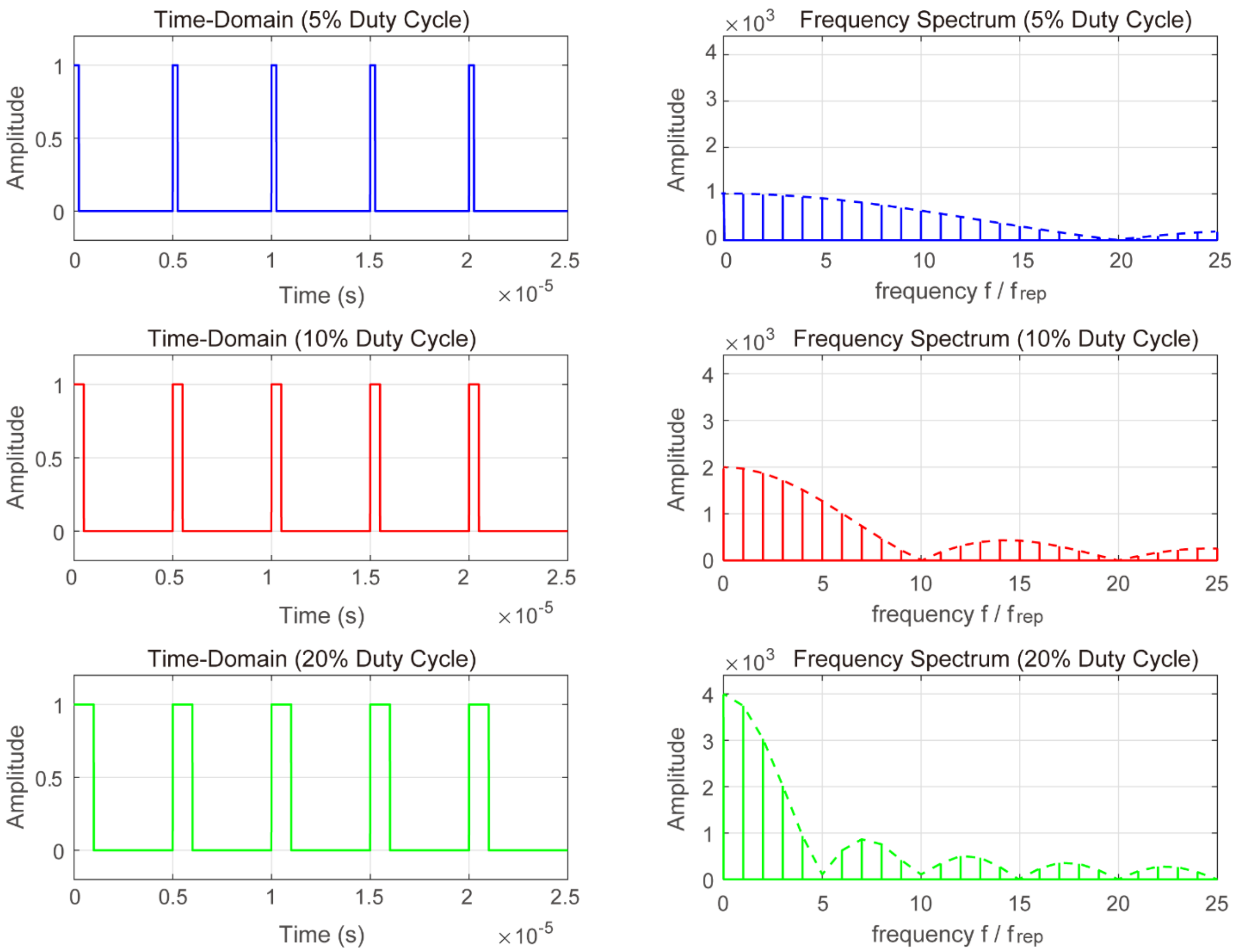

**Fig. S2:** Time-domain and frequency-domain simulation of pulses with different duty cycles.

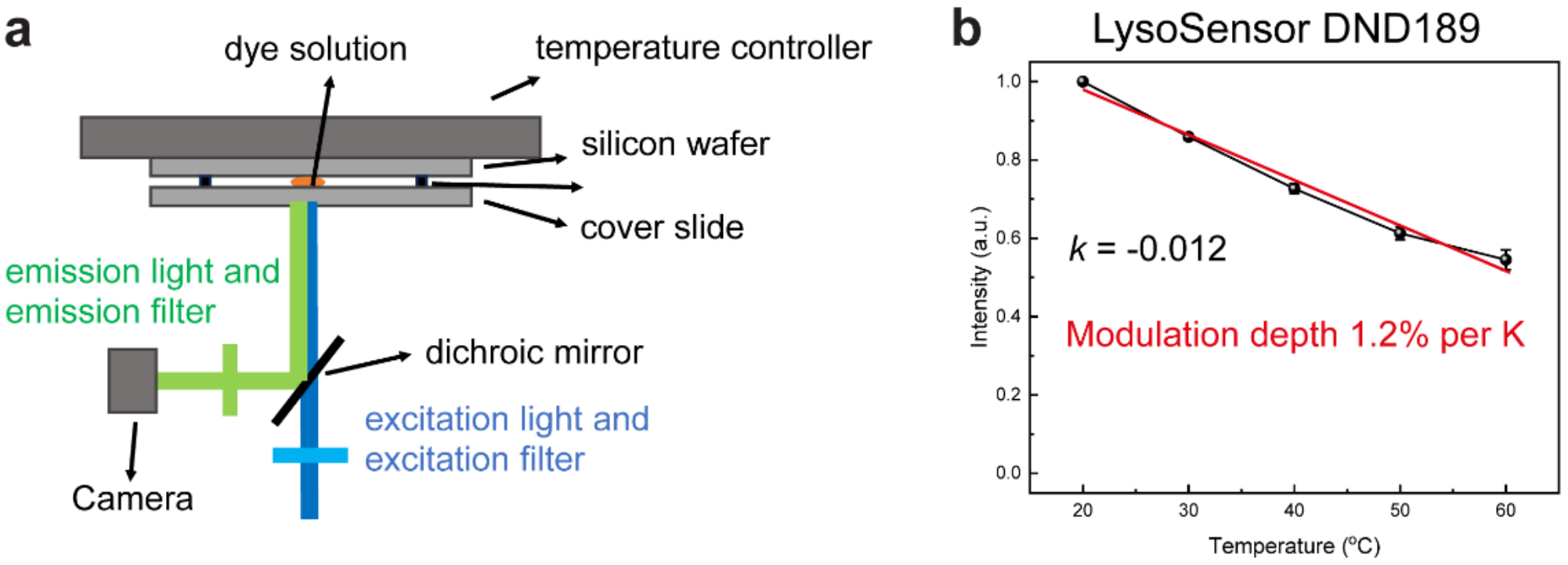

**Fig. S3: Fluorescence thermal sensitivity measurement.**

**a,** The system diagram of fluorescence thermal sensitivity measurement. **b,** Thermal sensitivity of LysoSensor DND189.

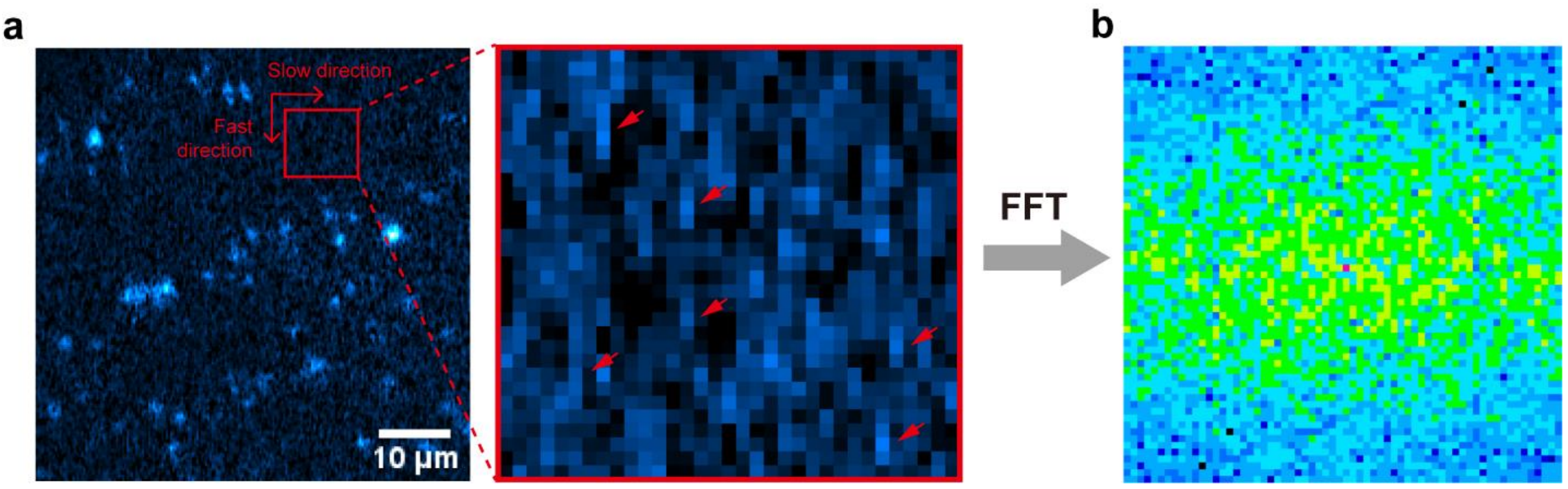

**Fig. S4: Noise correlation in FILM images.**

**a,** FILM image of lysosomes at 1711cm$^{-1}$. The red arrows indicate the correlation between pixels along the fast-scanning direction. **b,** Fourier spectrum of **a**. A clear decreasing trend from lower to higher frequencies along the fast axis suggests a structured noise pattern.

**Fig. S5:** FILM hyperspectral images with different IR frequency, highlighting the chemical selectivity.

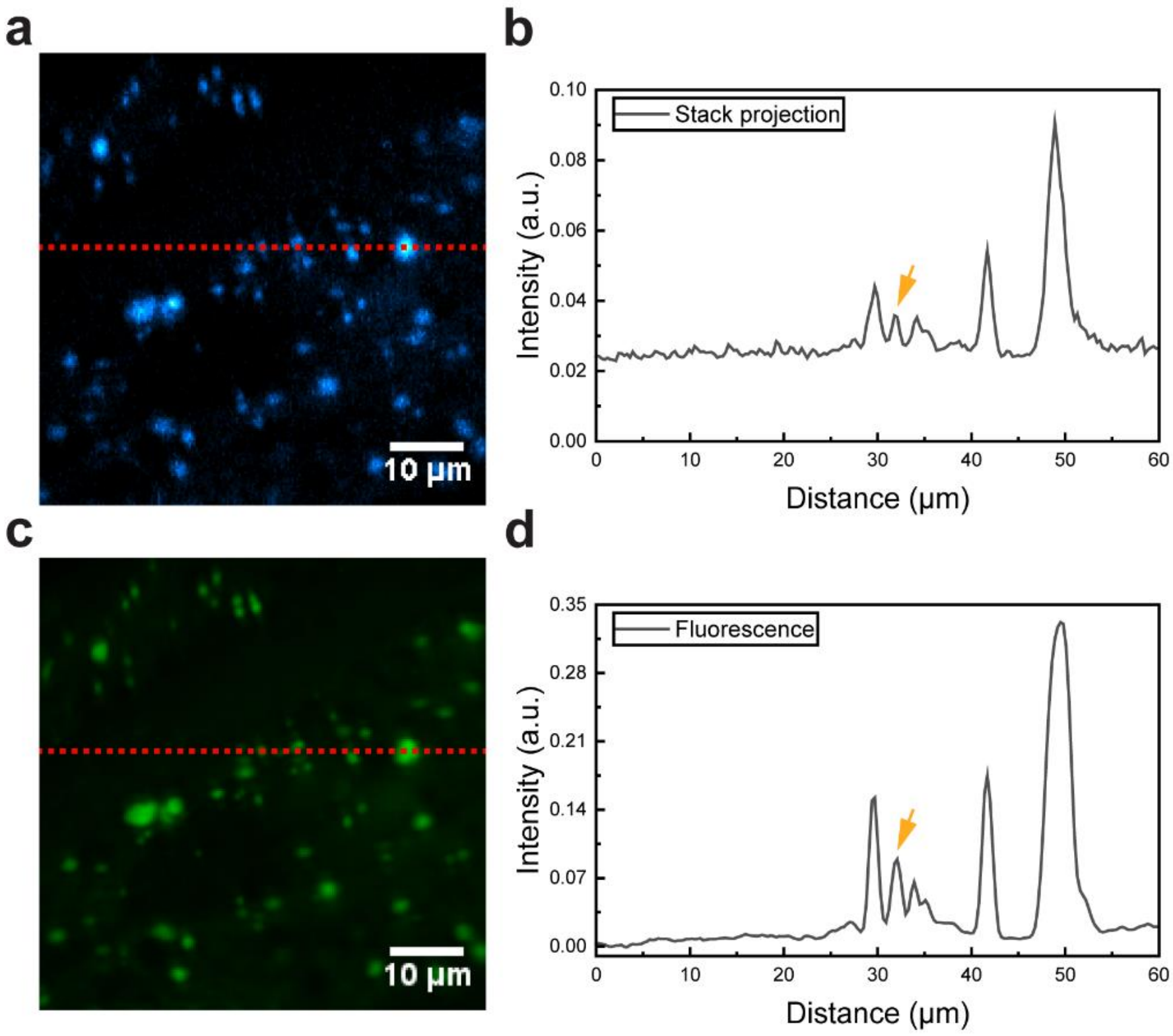

**Fig. S6: Spatial feature verification of SPEND denoising process.**

**a,** FILM hyperspectral imaging projection. **b,** Intensity profile across the red-dash line marked in **a**. **c,** DC fluorescence image. **d,** Intensity profile across the red-dash line marked in **c**.

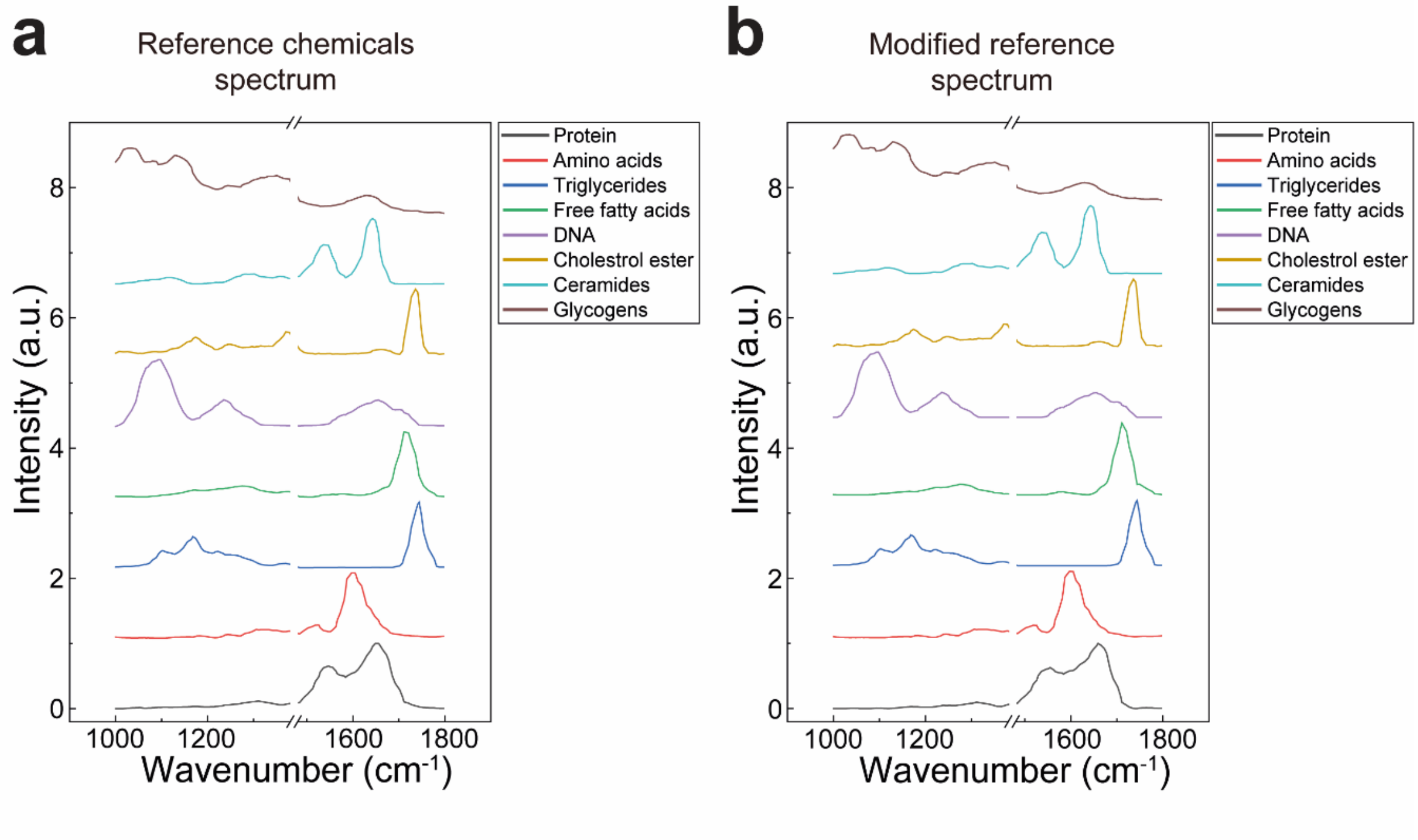

**Fig. S7: FILM spectrum of eight standards for spectral deconvolution.**

**a**, Reference spectrum of eight standards. **b**, MCR modified reference spectrum.

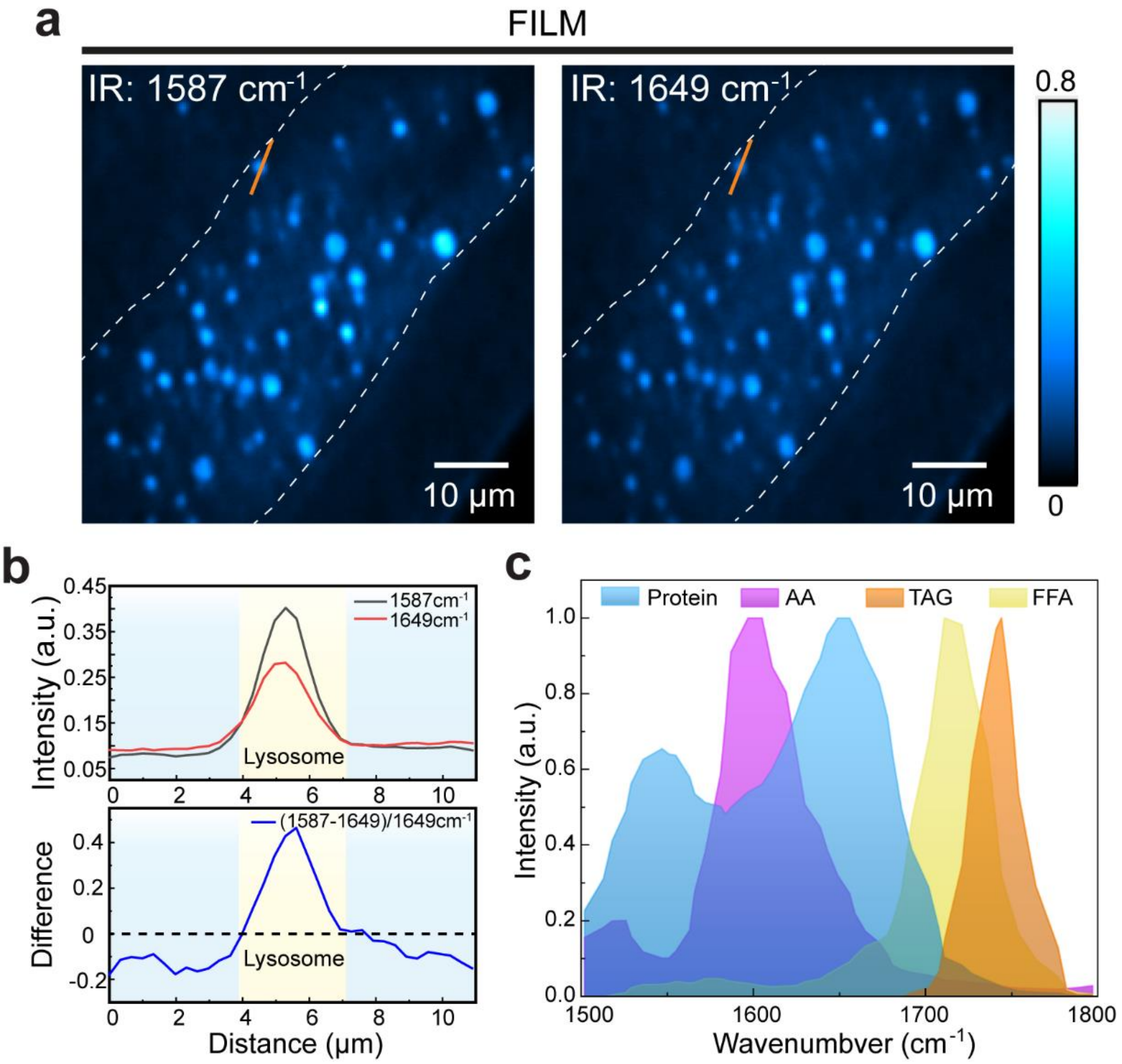

**Fig. S8: Lysosomes exhibit distinctive spectral features compared to the surrounding region.**

**a**, FILM images of *C. elegans* labelled with LysoSensor DND189. **b**, Intensity profiles along the orange lines marked in **a**. **c,** FILM spectrum of standard mixtures, including protein, amino acids (AA), triglycerides (TAG, lipid ester) and free fatty acids (FFA). These reference spectra were acquired with the same instrument and were used to assist in the assignment of characteristic peaks in lysosomal spectra. Scale bar: 10 µm.

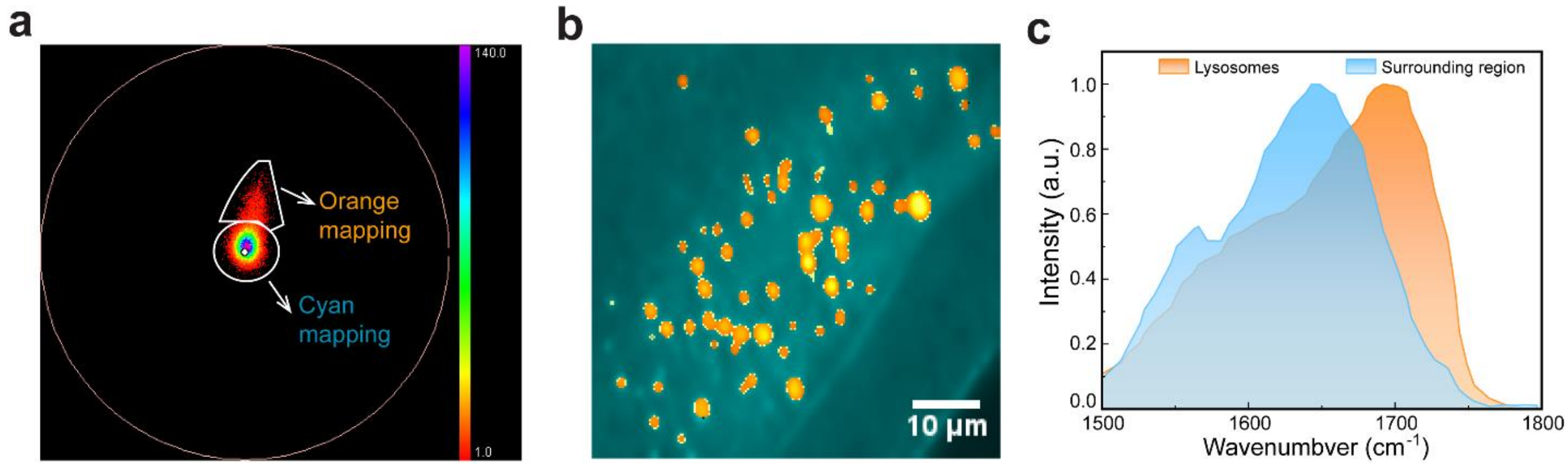

**Fig. S9: Spectral phasor analysis of FILM hyperspectral data.**

**a,** Spectral phasor plot on the hyperspectral image in **Supplementary Fig. 8**. **b,** Phasor segmentation map retrieved from the clusters marked in **a**. **c**, The spectral phasor analysis identified two distinct spectra in the FOV, corresponding to lysosomes and tissues, respectively.

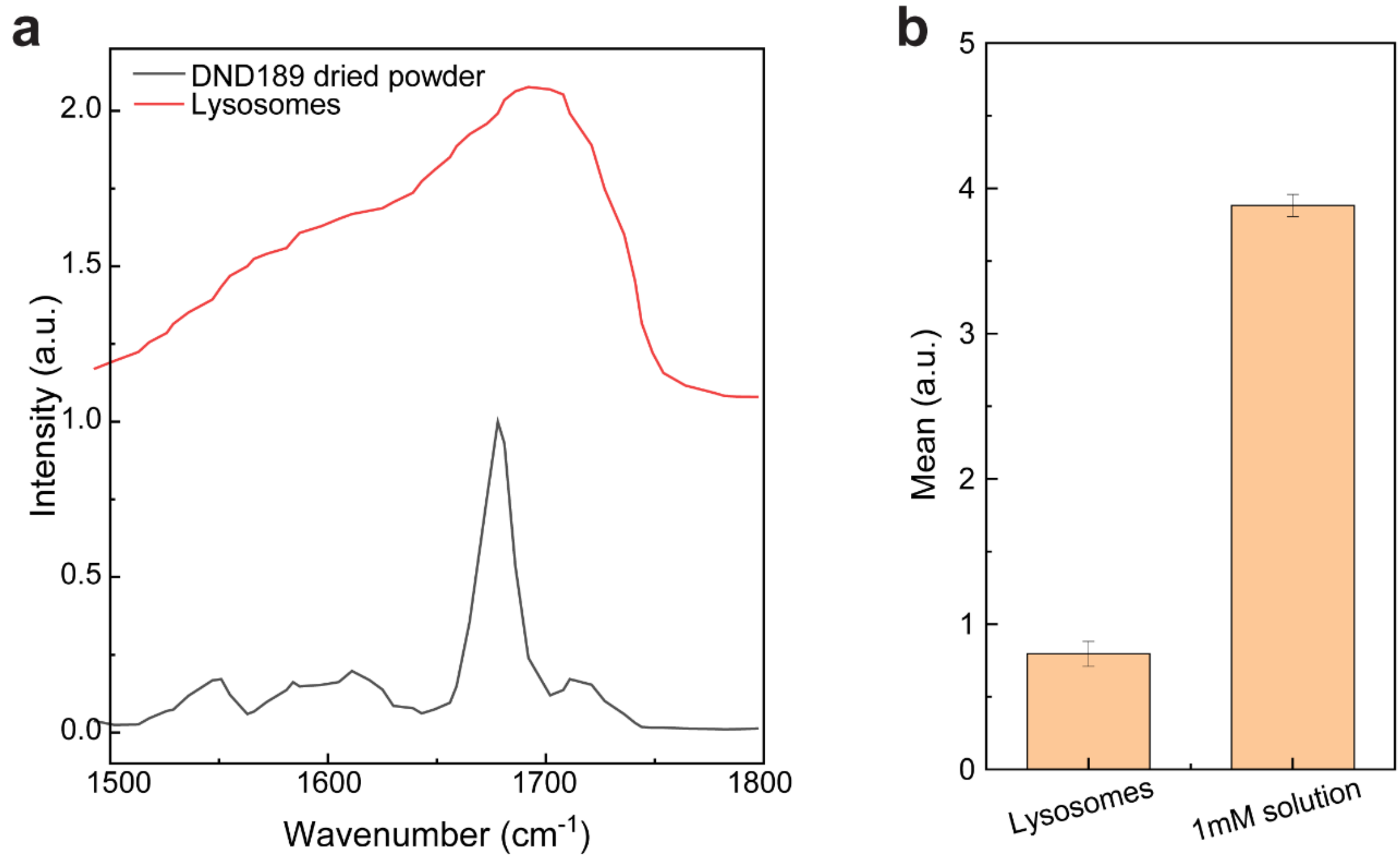

**Fig. S10: Evaluation of the influence of dyes on the spectrum.**

**a,** Comparison of lysosome spectra and LysoSensor DND189 dye. **b,** Fluorescence intensity comparison between lysosomes and standard 1 mM solution, indicating the dye concentration is below 1mM, which is lower than the concentration of the dominated biomolecules (n=8). Statistical data are presented as mean ± s.d.

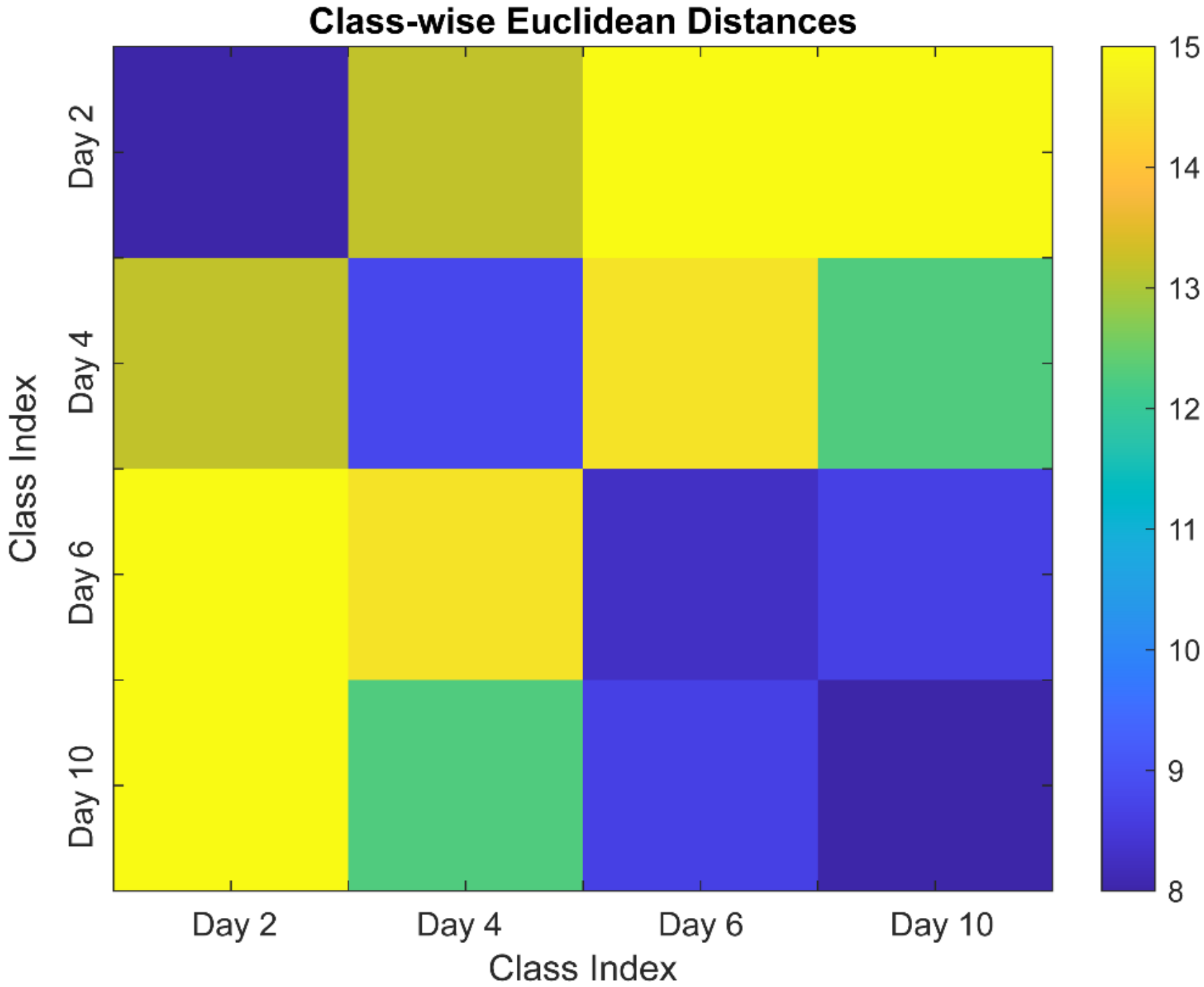

**Fig. S11:** Euclidean distance between data points calculated based on t-SNE and categorized into age groups.

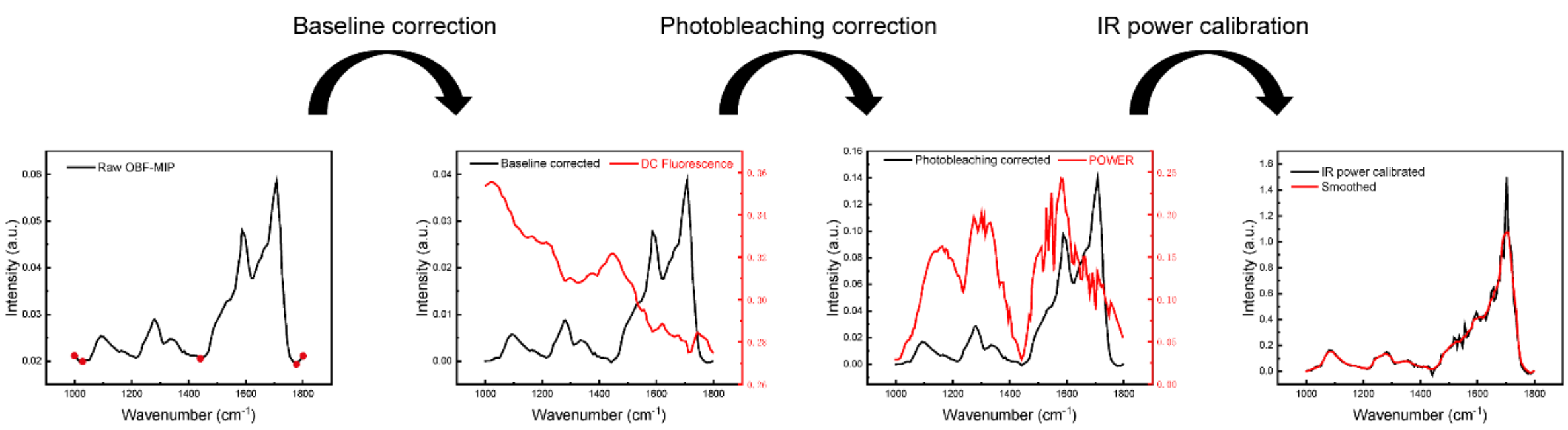

**Fig. S12: Data correction and calibration process for FILM spectra, including baseline correction, photobleaching correction, and IR power calibration.** The raw FILM spectrum (black) is baseline corrected, as indicated by the red dots, followed by photobleaching correction (red curve in the second panel), and then IR power calibration (red curve in the third panel) with smoothing applied. The red curve in the fourth panel represents the final output spectrum.

**Table S1:** Information on standards for spectral deconvolution

| Chemicals | Composition | Suppliers |
|---|---|---|
| Triglyceride mixture | Triacetin (C2:0), Tributyrin (C4:0), Tricaproin (C6:0), Tricaprylin (C8:0), Tricaprin (C10:0) | 17810, Supelco, Sigma-Aldrich |
| Protein mixture | Ribonuclease B, Insulin, Lysozyme, Transferrin, Bovine serum albumin (BSA), Trypsin inhibitor, β-lactoglobulin A, Carbonic anhydrase, Lactate dehydrogenase | MSRT2, Sigma-Aldrich |
| Ceramide mixture | L2-hydroxy C24:1, 2-hydroxy C18:1, C24:1, and C24:0 fatty acyl chains | 22853, Cayman Chemicals |
| Fatty acid mixture | Lauric Acid (C12:0), Myristic Acid (C14:0), Palmitic Acid (C16:0), Stearic Acid (C18:0), Arachidic Acid (C20:0), Lignoceric Acid (C24:0), Palmitoleic Acid (C16:1), Oleic Acid (C18:1), Nervonic Acid (C24:1) | 17942, Cayman Chemicals |
| DNA mixture | Sheared salmon sperm DNA solution, ~2000 bp average size | 15632011, Sigma-Aldrich |
| Amino acid mixture | Alanine, Arginine, Aspartic Acid, Cystine, Glutamic Acid, Leucine, Lysine, Serine, Threonine, Tyrosine, Valine, Histidine, Isoleucine, Methionine, Phenylalanine, Proline, Glycine, and Ammonium Chloride | AAS18, Supelco, Sigma-Aldrich |
| Cholesterol ester mixture | Cholesteryl Palmitate (C16:0), Cholesteryl Oleate (C18:1) | C6072, Sigma-Aldrich<br>C9253, Sigma-Aldrich |
| Glycogen | Glycogen (from bovine liver), highly branched glucose polymer | G0885, Sigma-Aldrich |

**Table S2:** IR marker peaks of eight references

| Substance | Wavenumber ($cm^{-1}$) | Assignments |
|---|---|---|
| Proteins | ~1650 | Amide I (C=O stretch; includes α-helix ~1650–1658, β-sheet 1618–1640/1680–1695, turn 1660–1680, random coil 1640–1650) |
| | ~1550 | Amide II (N–H bending + C–N stretching) |
| | ~1309 | Amide III (C–N stretch + N–H bend), α-helix–related region (~1310–1330) |
| Amino Acids | ~1590 | $\delta_{as}(NH_3^+)$ (asymmetric deformation); $\nu_{as}(COO^-)$ (asymmetric stretch) |
| | ~1520 | $\delta(NH_3^+)$ deformation (zwitterion) / aromatic ring modes |
| | ~1310 | ν(C–N) stretch with δ(NH) / CH bending (side-chain dependent) |
| Triglycerides | ~1744 | Ester C=O stretching (glycerol tri-ester) |
| | ~1230 | $\nu_{as}$(C–O–C) asymmetric stretch of ester |
| | ~1168 | $\nu_s$(C–O–C) symmetric stretch of ester |
| | ~1098 | ν(C–O) stretch of glycerol backbone / C–C skeletal vibrations |
| Free Fatty Acids | ~1711 | ν(C=O) of COOH |
| | ~1280 | ν(C–O) + δ(OH)_in-plane |
| | ~1220 | long-chain ν(C–C)/C–O combo & $CH_2$ wag/twist |
| DNA | ~1655 | Base vibrations (C=O in bases) |
| | ~1225 | $\nu_{as}(PO_2^-)$ asymmetric stretch (phosphate backbone) |
| | ~1080 | $\nu_s(PO_2^-)$ symmetric stretch (phosphate backbone) |
| Cholesterol Ester | ~1735 | Ester C=O stretching (cholesteryl ester linkage) |
| | ~1375 | $CH_3$ symmetric bending (acyl chains) |
| | ~1175 | $\nu_s$(C–O–C) symmetric stretch of ester |
| Glycogen | ~1340 | δ(C–H) deformation |
| | ~1130 | C–O–C asymmetric stretching |
| | ~1030 | C–O stretching (C–OH + C–O–C symmetric) + ring breathing |
| Ceramides | ~1640 | Amide I (C=O stretch of amide linkage) |
| | ~1540 | Amide II (N–H bending + C–N stretching) |
| | ~1285 | C–N stretching + N–H bending / $CH_2$ wagging |